\documentclass[twocolumn,superscriptaddress,amssymb,amsmath, showpacs,balancelastpage, nofootinbib]{revtex4-1}
\usepackage{graphicx,longtable,mathrsfs,color,array}
\usepackage[unicode=true,pdfusetitle,
bookmarks=true,bookmarksnumbered=true,bookmarksopen=true,bookmarksopenlevel=1,
breaklinks=false,pdfborder={0 0 0},backref=false,colorlinks=true]{hyperref}
\hypersetup{citecolor=blue,filecolor=blue,linkcolor=blue,urlcolor=blue,pdfauthor={Name}
}
\usepackage[usenames,dvipsnames]{xcolor}
\usepackage{amssymb,amsmath,mathtools,mathrsfs,enumitem}
\usepackage{epsfig,subfigure,placeins,float}
\usepackage{booktabs,longtable,multirow}
\usepackage{exscale,relsize}
\usepackage[normalem]{ulem}
\usepackage[T1]{fontenc}
\usepackage[utf8]{inputenc}
\usepackage{enumerate}
\usepackage{times, mathptmx}
\usepackage{tikz}
\usepackage{aas_macros}
\usetikzlibrary{arrows,positioning,decorations.markings,decorations.pathmorphing,calc}

\newcommand{\be}{\begin{equation}}
\newcommand{\ee}{\end{equation}}
\newcommand{\cgw}{c_{\rm GW}}
\newcommand{\MPl}{M_{\rm Pl}}
\newcommand{\Oo}{{{\cal O}(1)}}

\newcommand{\aT}{\alpha_T}

\newcommand{\comment}[1]{}

\newcolumntype{C}[1]{>{\centering\let\newline\\\arraybackslash\hspace{0pt}}m{#1}}

\def\Hz{{\rm Hz}}

\def\cgw{c_{\rm GW}}
\def\Ggw{G_{\rm GW}}

\def\L{{\cal L}}

\definecolor{hyperref}{RGB}{026,028,087}
\def\gsim{ \lower .75ex \hbox{$\sim$} \llap{\raise .27ex \hbox{$>$}} }
\def\lsim{ \lower .75ex \hbox{$\sim$} \llap{\raise .27ex \hbox{$<$}} }
\def\nn{\nonumber}
\def\ni{\noindent}

\newcommand*{\mathcolor}{}
\def\mathcolor#1#{\mathcoloraux{#1}}
\newcommand*{\mathcoloraux}[3]{%
\protect\leavevmode
\begingroup
\color#1{#2}#3%
\endgroup
}
\newlength{\stheight}
\newcommand\textst[1][fu-grey]{
\ifmmode\setlength{\stheight}{+1.0ex}
\else\setlength{\stheight}{+0.5ex}
\fi
\bgroup\markoverwith{\textcolor{#1}{\rule[\the\stheight]{2pt}{1.0pt}}}\ULon
} % strikethrough text command
\newcommand{\textins}[2][fu-grey]{
\ifmmode\mathcolor{#1}{#2}
\else\textcolor{#1}{#2}\@\,
\fi
}
\graphicspath{{./Plots/}}
\allowdisplaybreaks

\tikzstyle{vecArrow} = [thick, decoration={markings,mark=at position
1 with {\arrow[semithick]{open triangle 60}}},
double distance=1.4pt, shorten >= 5.5pt,
preaction = {decorate},
postaction = {draw,line width=1.4pt, white,shorten >= 4.5pt}]

\begin{document}

\setcounter{tocdepth}{5}
\title{Probing the speed of gravity with LVK, LISA, and joint observations}

\author{Ian Harry}
\email{Corresponding author: ian.harry@port.ac.uk}
\affiliation{Institute of Cosmology \& Gravitation, University of Portsmouth, Portsmouth, PO1 3FX, U.K.}

\author{Johannes Noller}
\email{Corresponding author: johannes.noller@port.ac.uk}
\affiliation{Institute of Cosmology \& Gravitation, University of Portsmouth, Portsmouth, PO1 3FX, U.K.}
\affiliation{DAMTP, University of Cambridge, Wilberforce Road, Cambridge CB3 0WA, U.K.}

\begin{abstract}
Theories of dark energy that affect the speed of gravitational waves $\cgw$ on cosmological scales naturally lead to a frequency-dependent transition of that speed close to the LIGO/Virgo/KAGRA (LVK) band. While observations such as GW170817 assure us that $\cgw$ is extremely close to the speed of light in the LVK band, a frequency-dependent transition below the LVK band is a smoking-gun signal for large classes of dynamical dark energy theories. Here we discuss 1) how the remnants of such a transition can be constrained with observations in the LVK band, 2) what signatures are associated with such a transition in the LISA band, and 3) how joint observations in the LVK and LISA bands allow us to place tight constraints on this transition and the underlying theories. 
We find that deviations of $\cgw$ can be constrained down to a level of $\sim 10^{-17}$ in the LVK {\it and} LISA bands even for mild frequency-dependence, much stronger than existing bounds for frequency-independent $\cgw \neq c$. We use the strain data from GW170817 to bound the deviation of $\cgw$ to be less than
$10^{-17}$ at 100 Hz and less than $10^{-18}$ at 500 Hz.
We also identify a particularly interesting type of transition in between the LVK and LISA bands and show how multi-band observations can constrain this further. Finally, we discuss what these current and forecasted constraints imply for the underlying dark energy theories.
\end{abstract}

\date{\today}
\maketitle

\tableofcontents

\section{Introduction} \label{sec-intro}
One of the most fundamental observables in testing our understanding of gravity is the speed with which gravity propagates. 
In a Lorentz-invariant solution, gravitational waves will propagate at the speed of light, but this need not be the case if Lorentz invariance is spontaneously broken, as it is in our Universe. 
Here, just as for any other waves, the speed of gravitational waves $\cgw$ depends on the medium they are propagating through. Measuring $\cgw$ therefore lets us probe the nature of this medium. 
With dark energy accounting for close to 70\% of the energy budget of our current Universe \cite{Planck:2018vyg}, constraining $\cgw$ provides an especially powerful tool in hunting for new degrees of freedom potentially associated with dark energy or a modified theory of gravity. Such new degrees of freedom, which non-trivially affect the background evolution and hence the universal `medium', are a generic consequence of departures from modelling dark energy as a cosmological constant $\Lambda{}$. Tight constraints found on some such theories \cite{Baker:2017hug,Creminelli:2017sry,Sakstein:2017xjx,Ezquiaga:2017ekz} -- also see earlier related work \cite{Amendola:2012ky,Amendola:2014wma,Deffayet:2010qz,Linder:2014fna,Raveri:2014eea,Saltas:2014dha,Lombriser:2015sxa,Lombriser:2016yzn,Jimenez:2015bwa,Bettoni:2016mij,Sawicki:2016klv} -- following the joint observations of GW170817 and GRB 170817A \cite{TheLIGOScientific:2017qsa,2041-8205-848-2-L14,2041-8205-848-2-L15,LIGOScientific:2017zic,LIGOScientific:2017ync} neatly illustrate this point. 

Current constraints on the speed of gravitational waves, $\cgw$, strongly depend on the frequency range in question and are conveniently expressed in terms of the parameter 
\begin{align} \label{aT_expression}
\aT \equiv (\cgw^2 - c^2)/c^2,
\end{align}
where $c$ is the speed of light. Note that $\aT$ therefore is a measure of the `tensor speed excess', i.e. it is positive for $\cgw > c$ and negative when $\cgw < c$.  We will remain agnostic here and will not assume anything about the sign of $\aT$ in what follows\footnote{It is useful to recall that for a spontaneously Lorentz-breaking solution, $\cgw \leq c$ need not be imposed by causality requirements -- in fact the opposite can be the case, see e.g. \cite{deRham:2019ctd,deRham:2021fpu} and references therein.}.
Constraints from the cosmic microwave background and large scale structure (see \cite{Noller:2018wyv,Bellini:2015xja,Hu:2013twa,Raveri:2014cka,Gleyzes:2015rua,Kreisch:2017uet,Zumalacarregui:2016pph,Alonso:2016suf,Arai:2017hxj,Frusciante:2018jzw,Reischke:2018ooh,Mancini:2018qtb,Noller:2018eht,Arai:2019zul,Brando:2019xbv,Arjona:2019rfn,Raveri:2019mxg,Perenon:2019dpc,Frusciante:2019xia,Arai:2019zul,SpurioMancini:2019rxy,Bonilla:2019mbm,Baker:2020apq,Joudaki:2020shz,Noller:2020lav,Noller:2020afd,Traykova:2021hbr}
and references therein), corresponding to frequencies $f \sim 10^{-18} - 10^{-14} \; \Hz$, require $|\aT| \lesssim {\cal O}(1)$. Binary pulsars, in particular the Hulse-Taylor binary, impose $|\aT| \lesssim {\cal O}(10^{-2})$ for frequencies $f \sim 10^{-4} \; \Hz$ \cite{Jimenez:2015bwa}. Gravitational wave observations of the afore-mentioned GW170817 event then place an extremely tight bound of  $|\aT| \lesssim {\cal O}(10^{-15})$ for frequencies $f \sim 10^1 - 10^{3} \; \Hz$  \cite{TheLIGOScientific:2017qsa,2041-8205-848-2-L14,2041-8205-848-2-L15,LIGOScientific:2017zic,LIGOScientific:2017ync} , while the presence of high energy cosmic rays (and inferred absence of gravitational Cherenkov radiation) leads to a $\aT \gtrsim -{\cal O}(10^{-15})$ bound at energies of $\sim 10^{10}$ GeV \cite{Moore:2001bv}, i.e. for very high frequencies $f \sim 10^{34} \; \Hz$. Note that the last bound {\it does} depend on the sign of $\aT$. Finally, several forecasted bounds exist for future LISA observations \cite{Littenberg:2019mob,Baker:2022rhh}. Since these are particularly relevant in the context of this paper, we will discuss them in more detail in section \ref{sec-probe-LISA}.
The frequency-dependence of the bounds on $\aT$ is particularly important, since the theoretical predictions for this parameter are also frequency-dependent. More specifically, for large classes of dark-energy related models that yield a significantly non-zero $\aT$ on cosmological scales, i.e. at very low frequencies, one naturally expects a transition back towards $\aT = 0$ close to or somewhat below $10^2 \; \Hz$ \cite{deRham:2018red}. This means that the tight GW170817 constraint on $\aT$ is perfectly consistent with models that yield a detectable, non-zero $\aT$ at lower frequencies.

Current ground-based observatories, Advanced LIGO~\cite{LIGOScientific:2014pky}, Advanced Virgo~\cite{VIRGO:2014yos} and KAGRA~\cite{Aso:2013eba} have sensitivity between approximately $20\; \Hz$ and $\sim 2000 \; \Hz$.
In the coming years, these observatories will continue to run a series of
increasingly sensitive observing runs, with lower frequency sensitivity
reaching no lower than $10\; \Hz$~\cite{KAGRA:2013rdx}.
It is expected that these runs will yield numerous observations of binary neutron star mergers, and hopefully numerous multimessenger observations~\cite{KAGRA:2013rdx}, with which powerful tests of the speed of gravity can be performed.
However, to perform tests outside of this frequency range we must wait for the next generation of gravitational-wave observatories, expected to become operational in the 2030s.
The third generation of ground-based observatories, Einstein Telescope~\cite{Punturo:2010zza} and Cosmic Explorer~\cite{Reitze:2019iox}, will increase overall sensitivity, and increase the range of sensitive frequencies, covering from $\sim 1\; \Hz$ up to $\sim 10^4\; \Hz$.
Perhaps of most interest in this context though are the proposed space-based observatories, LISA~\cite{LISA:2017pwj} and TianQin~\cite{TianQin:2015yph}, which will cover a range of frequencies inaccessible from the ground, covering $\sim10^{-4}\; \Hz$ to $\sim10^{-1}\; \Hz$.

In this paper we investigate how observations from the LIGO/Virgo/KAGRA observatories (henceforth LVK), and GW170817 in particular, can be
leveraged to provide additional constraints on a frequency-dependent
transition of $\cgw$, extending existing constraints that assume a frequency-independent $\cgw$. We also explore how one can use lower frequency measurements of gravitational waves in the LISA band to test for the existence and nature of such a transition and how one might be able to combine information from both space-based and ground-based observation bands to provide further
constraints.

The paper is structured as follows. In section \ref{sec-freq} we discuss in what sense the frequency-dependence of $\cgw$ is a generic phenomenon in models that do lead to a non-zero $\aT$ on cosmological scales. As a key outcome of this discussion, we introduce and motivate templates to capture this frequency dependence. In section   
\ref{sec-signal} we investigate what the imprint of a frequency-dependent $\cgw$ is on the gravitational wave signal observed, highlighting the resulting stretching and squeezing 
of the waveform as the main associated observable. In the following sections we then compute constraints and forecasts on the frequency-dependence of $\cgw$: In section \ref{sec-probe-LIGO} we present constraints computed for the LVK band, that extend existing bounds on a constant and hence frequency-independent $\cgw$. In section \ref{sec-probe-LISA} we present analogous forecasts for the LISA band, and in section \ref{sec-probe-joint} we discuss how joint observations in both bands will allow us to place highly precise constraints on $\cgw$ over the whole range of relevant frequency and energy scales. We conclude and summarise our findings in \ref{sec-conc} and collect additional details in the appendices.

\section{From theory to templates} \label{sec-freq}

In this section we will illustrate how a non-luminal $\cgw$ can arise in theories of dark energy (\ref{subsec-decgw}), why this generically leads to a $\cgw$ that is frequency-dependent, where (if indeed present) one would expect this to be especially pronounced close to the frequencies observed by present (LVK) and near-future (LISA/TianQin/Einstein Telescope/Cosmic Explorer) detectors (\ref{subsec-reach}), and how one can use this information to build theoretically well-motivated templates to be used in constraining and searching for frequency-dependent $\cgw$ (\ref{subsec-template}).

\subsection{Dark energy and \texorpdfstring{$\cgw \neq c$}{[cgw neq c]}} \label{subsec-decgw}

Since GR is the single consistent theory of a massless spin-2 field, testing for (potentially dark energy-related) deviations away from it generically amounts to probing the presence of new gravitational degrees of freedom. Scalar-tensor theories are a minimal deviation from GR in this sense, as they only introduce a single additional degree of freedom. Accordingly, Horndeski gravity \cite{Horndeski:1974wa,Deffayet:2011gz}\footnote{For the equivalence between the formulations of \cite{Horndeski:1974wa} and \cite{Deffayet:2011gz}, see \cite{Kobayashi:2011nu}.}, the most general Lorentz-invariant scalar-tensor action that gives rise to second order equations of motion, has recently been the main workhorse in testing for deviations from GR. In appendix \ref{app-horn} we will summarise the key relevant results for the complete Horndeski theory, but here we will illustrate the salient points with the following example scalar-tensor theory\footnote{We emphasise that we merely use \eqref{example-lag} as an example to illustrate generic $\aT$ features in scalar-tensor theories. While they will not be important here, we refer the reader curious about some of the interesting stability/screening/positivity properties of \eqref{example-lag} to \cite{Kobayashi:2011nu,Bellini:2014fua,Berezhiani:2013dw,Koyama:2013paa,Melville:2019wyy} and references therein.} 
\begin{align} \label{example-lag}
{\cal L} = \frac{\MPl^2}{2}R -\frac{1}{2}\nabla_\mu \phi \nabla^\mu \phi + \frac{\MPl^2}{\Lambda^3} g(\phi)G_{\mu\nu}\nabla^\mu\nabla^\nu\phi.
\end{align}
The first two terms are the standard kinetic interactions for the massless tensor $g_{\mu\nu}$ and the scalar $\phi$, respectively, while the last term is a higher-order interaction involving both the scalar and tensor. $G_{\mu\nu}$ is the Einstein tensor and $g(\phi)$ is a dimensionless function of $\phi/\MPl$. 
There are two key mass scales: the Planck mass $\MPl$, and the scale $\Lambda$ associated with second derivatives of the scalar. In cosmology these are linked to the value of the Hubble constant today, $H_0$, and are conventionally taken to satisfy  $\Lambda^3 = \MPl H_0^2$. This choice ensures that all interactions can give $\Oo$ contributions to the (cosmological) background evolution, so is phenomenologically motivated in order to allow the scalar $\phi$ to act as dark energy.

We can now work out the speed of gravitational waves $\cgw$ for the theory \eqref{example-lag}. 
In a scalar-tensor theory as we are considering here, gravitational waves propagate through a `medium' provided by the presence of the scalar $\phi$. The interactions of $\phi$ with $g_{\mu\nu}$ encoded in \eqref{example-lag} can affect the refractive index of this medium and hence $\cgw$. More specifically, we find
\begin{align}
\alpha_{T} &=\frac{2(\nabla\bar\phi)^2 g'(\bar\phi)}{\MPl\Lambda^3-(\nabla\bar\phi)^2 g'(\bar\phi)},
\label{example-aT}
\end{align}
where a prime denotes a derivative with respect to the argument of the function (we recall this is $\phi/\MPl$ for $g$ and so both $g$ and $g'$ are dimensionless)
and $\bar \phi$ is the background solution for the scalar. This makes it clear that for a Lorentz-invariant background solution with $\bar\phi = 0$ we obtain $\aT = 0$ and hence gravitational waves travel at the speed of light there, as expected. However, when considering the cosmological background solution $\bar\phi = \bar\phi(t)$ of interest here, this time-dependent scalar evolution spontaneously breaks Lorentz invariance and provides a non-trivial medium for gravitational waves to travel through.
In this case, $\aT$ will generically be non-zero and we obtain the following dispersion relation
\begin{align} \label{example-dispersion}
\omega^2 = \cgw^2(t)k^2,
\end{align}
where $\cgw^2 = (1+\aT)c^2$ as above. We emphasise that this means $\aT$ is generically {\it time-dependent}, but (so far) {\it frequency-independent} -- we will see how frequency-dependence enters below. 
Regarding the time-dependence, note that the characteristic timescale for this in theories of dark energy is a Hubble time ($1/H_0 \sim 10^{10}$ years). 
Compared to the characteristic travel time for the signals seen by gravitational wave detectors this is approx. the same order of magnitude for a source at a distance of $\sim$ Gpc (travel time $\sim 3\times10^{9}$ years) and about two orders of magnitude larger for closer sources (the travel time is $\sim 10^8$ years for a source at $40$ Mpc), assuming the signal travels at close to the speed of light.
While taking into account this time-dependence can therefore have a noticeable effect for far-away sources, e.g.  modulating some of the relevant observables such as arrival times, this will not change the qualitative, order-of-magnitude constraints we are focusing on here.
We will therefore ignore any time-dependence here and leave an exploration of this effect to future work\footnote{Note that, while dark energy only becomes a major driver of the cosmological evolution from $z \sim {\cal O}(1)$ onwards, it only takes light approx. twice as much time to travel to us from redshift 1 vs. from redshift 1000, so this difference is immaterial to what we are doing here.}.

\subsection{The reach of a theory and frequency-dependent \texorpdfstring{$\cgw$}{[cgw]}} \label{subsec-reach}

Let us return to the example theory \eqref{example-lag}. The higher-order interactions encoded in the final term mean this theory is predictive at most up to the energy scale $\Lambda$ and cannot resolve energy scales close to or larger than this -- see e.g. \cite{Georgi:1993mps,Donoghue:1995cz,Burgess:2007pt} for reviews of the underlying field theoretic reasoning. This is precisely analogous to what happens in GR, which also encodes higher-order interactions within the Einstein-Hilbert term and where predictivity is lost as one approaches the corresponding energy scale there, namely the Planck scale. For cosmologically motivated models, where the scalar $\phi$ is linked to dark energy, we recall that $\Lambda^3 = \MPl H_0^2$. Expressed as a frequency this amounts to 
\begin{align}
\label{LambdaCutoff}
\Lambda \sim 260 \; {\rm Hz}.
\end{align}
This is the largest possible energy/frequency scale, a so-called cutoff, where \eqref{example-lag} stops being applicable. Making predictions around or above this scale then requires knowledge of a more complete high energy theory, a UV completion, that extends the regime of validity of this theory beyond the frequency/energy scale $\Lambda$. In the absence of knowledge about such a UV completion, i.e. in the field's present situation, we simply do not know how to relate a measurement at energies $\gtrsim \Lambda$ to a theory such as \eqref{example-lag}. Note that this conclusion does not rely on the specific form of \eqref{example-lag} and affects all sectors of the theory involving $\phi$ interactions\footnote{More specifically, it is precisely the interactions that affect $\aT$ which depend on $\Lambda$ and hence limit the applicability of the theory to energy scales below $\Lambda$. In other words, if $\phi$ interactions in a Horndeski theory contribute to the dark energy evolution today and affect $\aT$ on cosmological scales, this naturally leads to a cutoff \eqref{LambdaCutoff}. So this conclusion does indeed not rely on the specific example \eqref{example-lag}. Some (but not all) well-known  dark energy theories that do not affect $\aT$ can have a much larger cutoff, e.g. quintessence \cite{Ratra:1987rm,Wetterich:1987fm,Ferreira:1997hj,Caldwell:1997ii} or k-essence \cite{ArmendarizPicon:1999rj,Garriga:1999vw,ArmendarizPicon:2000dh,ArmendarizPicon:2000ah} theories.}. It is worth emphasizing that $\Lambda$ is the {\it largest} possible scale up to which such a theory 
can be predictive. Where precisely this threshold is depends on the specific nature of the UV completion. Compare this with the Fermi theory for weak interactions, where the naive cutoff scale is $\sim$ TeV, yet new physics associated with a UV completion already enters an order of magnitude below at the scale of the W-boson at $\sim 80$ GeV. The LISA and lower end of the LVK bands (as well as intermediate bands) are therefore particularly well-motivated regions to look for the onset of new physics associated with UV completions of dark energy theories.

How is the above discussion of regimes of validity for scalar-tensor theories relevant to the main observable we are considering in this paper, $\cgw$? This is discussed in detail in \cite{deRham:2018red} and we here summarise the essential results relevant for this paper. When deriving \eqref{example-aT} and \eqref{example-dispersion} we were implicitly working on cosmological scales, firmly within the regime of validity of \eqref{example-lag}. However, as one approaches the cutoff of the theory $\lesssim \Lambda$, the existence of a UV completion will manifest itself by additional interactions (highly suppressed on cosmological scales) becoming important. Precisely what these interactions are is determined by the (unknown) UV completion, but we can
infer some types of candidate additional interactions from investigating radiative corrections to generic Horndeski theories \cite{Luty:2003vm,Nicolis:2004qq,deRham:2010eu,Burrage:2010cu,deRham:2014wfa, Pirtskhalava:2015nla,deRham:2012ew,Goon:2016ihr,Saltas:2016nkg,Noller:2018eht,Heisenberg:2019wjv,Heisenberg:2020cyi,Goon:2020myi}. 
The example set of such interactions focused on in \cite{deRham:2018red} is
\begin{align}
\L^{(n)}=\MPl^2 G_{\mu\nu} \frac{\Box^n}{M^{2n+4}}(\nabla^\mu\phi\nabla^\nu\phi),
\label{Ln}
\end{align}
where $M \lesssim \Lambda$ is the scale associated with these UV-induced new operators and $n\geq 2$. Of particular relevance for our context is the (representative and generic) higher-derivative nature of these interactions. Including these interactions modifies the dispersion relation \eqref{example-dispersion} and one now finds \cite{deRham:2018red}
\begin{align} \label{dispersion}
\omega^2 \sim \cgw^2 (0) k^2 +\sum_{n\ge 2}\frac{c_n \Lambda^6}{3M^{2n+4}}(c^2 k^2-\omega^2)^{n-1}(\omega^2 + {\cal O}(k^2)),
\end{align}
where $\omega$ is the angular frequency, $k$  the wavenumber, the $c_n$ are some order one coefficients (whose precise value is not important here) and we have ignored suppressed $H$-dependent corrections, since $H^2/\omega^2 \ll 1$ for the frequencies relevant here. Here we explicitly write $\cgw$ as a function of $k$ and hence denote the asymptotic value of $\cgw$ for low energies/frequencies, i.e. effectively its cosmological value derived in \eqref{example-aT}, by $\cgw (0)$ -- see \cite{deRham:2018red} for details on the derivation of the full dispersion relation for this example as well as other related examples. 
Crucially, \eqref{dispersion} asymptotes to the cosmological $\cgw = \cgw (0)$ for small $k$ (low frequencies) and asymptotes to luminal $\cgw = c$ for large $k$ (high frequencies), where the $(c^2 k^2-\omega^2)^{n-1}$ drives this large $k$ approach to luminality. Here it is important to emphasise the point made by \cite{deRham:2018red} that asymptoting to $\cgw = c$ at high energies, as is entailed by \eqref{dispersion}, is a generic consequence of any UV completion as long as Lorentz invariance is restored at scales $\gtrsim M$. So this feature should certainly not be seen as an accident of the example interactions shown here.

Finally, it is instructive to consider the asymptotic scaling of $\cgw$ with $k$ deriving from \eqref{dispersion}. We focus on a single $n$ interaction in \eqref{Ln} and denote the values of $k$ and $f$ corresponding to the transition scale $M \lesssim \Lambda$ by $k_\star$ and $f_\star$, respectively, in what follows. Doing so, we find
\begin{align}
&k\ll k_\star:  &\cgw - \cgw(0) &\propto k^{2n-2}, \nn \\
&k\gg k_\star:  &c - \cgw &\propto k^{-2}.
\end{align}
We see that there is a clean power law scaling in both limits, where the relevant power for small $k$ depends on the precise interaction term present, while there is a universal $1/k^2$ approach to luminal $\cgw$ for large $k$. With an eye on constructing templates below, it is very important to separate what is a generic feature any template should recover vs. what is an accident of the example chosen. At scales below the transition scale $k_\star$, a UV completion will indeed generically give rise to higher-derivative interactions such as \eqref{Ln}. In the dispersion relation these will manifest themselves as higher powers of $k$ suppressed by the scale $M$, so generically one does indeed expect that a power law scaling is an excellent approximation for the low energy approach into the $\cgw$ transition. However, once reaching energy scales $\sim M$ (or equivalently $k_\star$) no firm prediction of the transition behaviour can be made without the (unknown) UV completion. In particular, the transition can in principle happen arbitrarily quickly and need not even follow a power law. So the universal $1/k^2$ scaling found for the simple example above should {\it not} be hard-coded into any template\footnote{As we shall see in more detail later, transitions that do have a `slow' $1/k^2$ scaling at high energies are severely constrained by LVK measurements such as GW170817 even when the analogue of $f_\star$ (and hence $k_\star$) are firmly within the LISA band. But while this is of course an interesting finding for transitions with this scaling in its own right, it should not be mistaken for a generic conclusion.}. Furthermore, when going beyond the simple example above and considering a generic situation where all $c_n$ can be non-zero, a different large $k$ scaling can be obtained for a given $\cgw(0)$ by carefully tuning the $c_n$. Judging whether such a tuning is natural again requires detailed knowledge of the UV completion, but this already provides further reason why no fixed power-law dependence for large $k$ should be hard-coded into any template\footnote{We thank Scott Melville for related discussions.}.

\subsection{Templates for \texorpdfstring{$\cgw$}{[cgw]}} \label{subsec-template}

In the sections below we will use current/forecasted observational data from LVK and LISA bands to place constraints on the frequency-dependence of $\cgw$. Doing so generically requires a template for the waveform and, in particular, for $\cgw$ itself. Given the above considerations and plethora of potentially contributing interactions, the precise form of $\cgw(f)$ in principle depends on a large number of parameters. This includes the $c_n$ in \eqref{dispersion}, and crucially parameters associated to the unknown UV completion that generically becomes important in or close to the LVK/LISA bands for dark energy theories leading to a non-luminal $\cgw$ at very low frequencies, i.e. for cosmology.

We would therefore like to work with a template for $\cgw$ that, while simplifying the parametric dependence and making an analysis practical, still captures the salient features of this frequency-dependence. 
As discussed above, such a template ought to 
asymptote to a luminal $\cgw$ at large frequencies (to enable consistency with LVK bounds), asymptote to a constant $\cgw (0)$ at small frequencies (consistent with \eqref{example-aT}), while also allowing different scalings with $k$ throughout the transition. A minimal template will depend on at least three parameters
\begin{itemize}
\item A speed $\cgw (0)$, which corresponds to the asymptotic speed of gravitational waves at low frequencies. Effectively this is the cosmological $\cgw$ that can differ significantly from $c$.
\item A frequency scale $f_\star$, denoting the `central frequency' of the transition from $\cgw (0)$ to $c$. In terms of the underlying physics this is set by the scale where new physics associated with the UV completion becomes important, e.g. $M$ in \eqref{Ln}.
\item A parameter $\sigma$, which controls how quickly the transition takes place. This is effectively a measure of the nature of the new interactions present due to the UV completion. 
\end{itemize}
Note that there can in principle be many more parameters controlling interactions that determine how quickly the transition takes place, so introducing a single parameter for this should be seen as a lowest order approximation. 

\begin{figure}[t!]
\centering
\includegraphics[width=\linewidth]{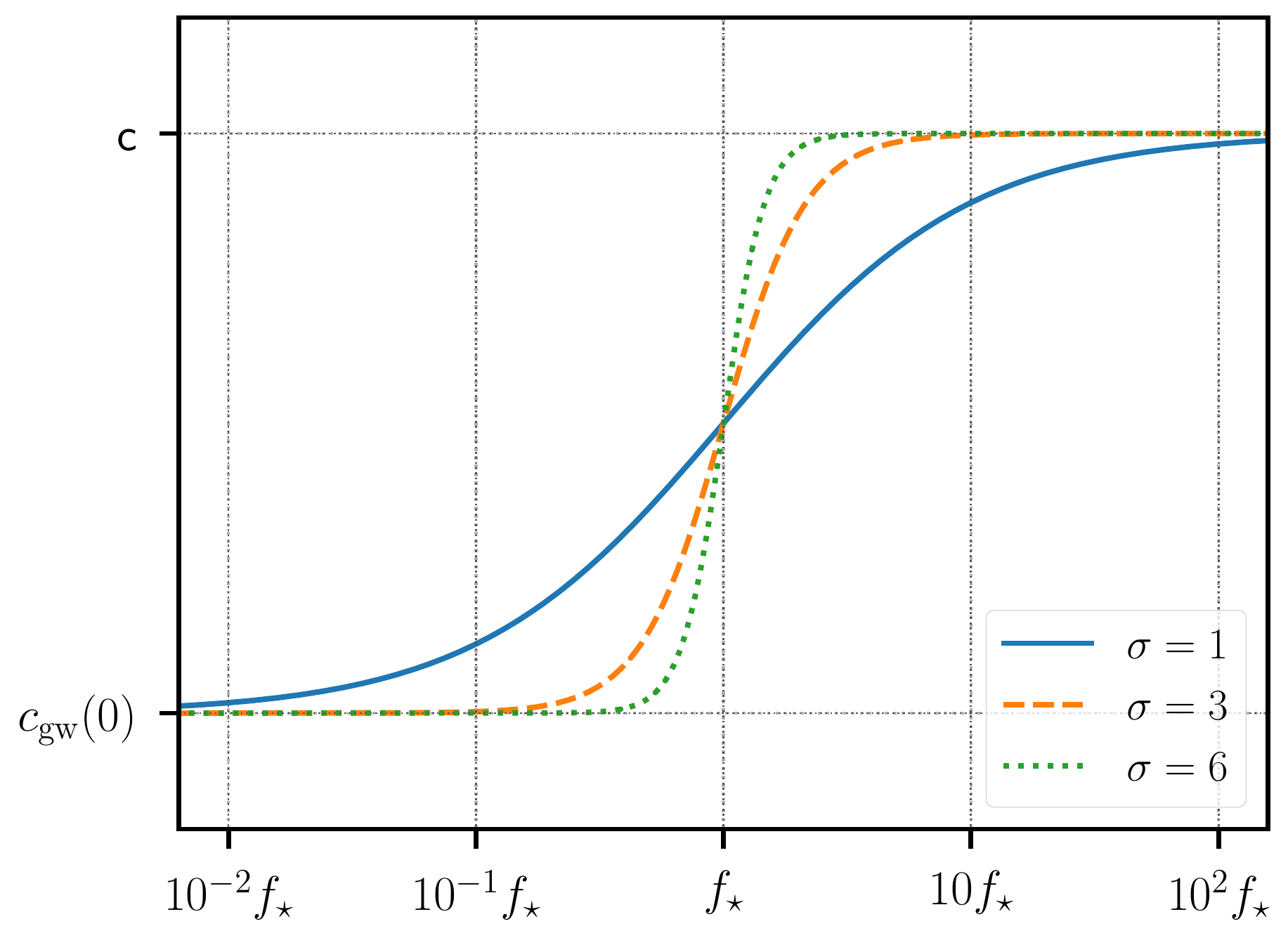}
\caption{
The speed of gravitational waves, $\cgw$, as a function of frequency $f$ for the templates discussed in section \ref{sec-freq}. The different curves correspond to different choices of the `steepness' parameter $\sigma$ that also controls the asymptotic power law scalings of the $\cgw$ -- see \eqref{sTemp_powerlaw}. $\cgw$ neatly interpolates between $\cgw(0)$, the speed of GWs on cosmological scales, and the speed of light $c$ at large frequencies. $f_\star$ physically gets set by the scale $M \lesssim \Lambda$ that marks the onset of the UV completion for the dark energy theory in question, cf. \eqref{Ln}.}
\label{fig-cgwEvo}
\end{figure}

Using the parameters listed above, we can build the following useful hyperbolic fitting function
\begin{align}
\Theta_\pm(f,\sigma,f_\star) &\equiv \tfrac{1}{2} \pm \tfrac{1}{2}\tanh\left[\sigma \cdot \log\left(f/f_\star\right)\right],
\label{Thetas}
\end{align}
where $\Theta_+$ transitions between 0 and 1 around $f_\star$, while $\Theta_-$ transitions from 1 to 0 around the same frequency. $\sigma$ controls how quickly this transition takes place, as advertised. With this function we can now build a straightforward template for $\cgw$ as a function of frequency $f$, that captures the essential features outlined above
\begin{align}
\label{GW_speed}
\cgw(f, \sigma, f_\star) = \cgw^{(0)} + (c - \cgw^{(0)})\Theta_+.
\end{align} 
This behaviour is illustrated in Figure~\ref{fig-cgwEvo}.

It is again instructive to consider asymptotic scalings with $k$. From \eqref{GW_speed} we find
\begin{align}
&k \ll k_\star:  &\cgw - \cgw(0) &\propto k^{2\sigma}, \nn \\
&k \gg k_\star:  &c - \cgw &\propto k^{-2\sigma}.
\label{sTemp_powerlaw}
\end{align}
This clearly shows that the steepness parameter $\sigma$ directly controls the power law scaling in both asymptotic limits, allowing us to mimic different UV-completion-induced scalings at high energies. Due to the simplicity of the template, the single parameter $\sigma$ controls both asymptotes and hence we have symmetric $k^{2\sigma}$ and $k^{-2\sigma}$ scalings for this template. There is no fundamental reason why the transition ought to be symmetric in this way and it is straightforward to refine the template to include asymmetric transitions upon introducing an additional parameter -- see appendix \ref{app-templates}\footnote{We also emphasise that the most highly UV sensitive (and hence least robust) features of the transition, e.g. the precise functional form around $f_\star$ in \eqref{dispersion} (which are generically different for the small and large frequency asymptote) are not reproduced by the template by design.}.
However, for the constraint analysis in the following sections we will work with the minimal template \eqref{GW_speed} and leave a more detailed analysis of more refined templates including additional corrections for future work. 

\section{The gravitational wave signal} \label{sec-signal}

Before proceeding to probing $\cgw$ using current and forecasted data, we will find it useful to understand some key features of the GW signal itself.

\subsection{Waveform stretching or squeezing}
If any detectable form of frequency-dependence (we will be more quantitative below) takes place within the observable LVK and/or LISA bands, where the source is tens to thousands of megaparsecs away from the detectors, the leading-order effect of a frequency-dependent $\cgw$ will be that the observed signal is stretched (for negative $\alpha_T$) or squeezed (for positive $\alpha_T$) as the travel time varies with frequency.

The travel time of a signal between a source and an observer on the Earth can be expressed as
\begin{align}
t = \frac{D_L}{\cgw(f)} \equiv \frac{D_L}{c ( 1 - \delta \cgw(f))} \approx \frac{D_L}{c} + \frac{D_L}{c} \delta \cgw(f),
\end{align}
where ${D_L}$ is the luminosity distance to the source. We define here
$\delta \cgw(f) \equiv 1 - \cgw(f) / c$, which is the fractional variation in the gravitational-wave speed from $c$. This is related
to $\aT$ according to $\aT = -2\delta \cgw + (\delta \cgw)^2$, the latter term being negligible in the cases that we will consider.
$\frac{{D_L}}{c} \delta \cgw(f)$ then denotes the frequency dependent time delay that is incurred if $\cgw$ is less than $c$.
As the frequency of radiation emitted by the source increases, waves emitted at different times (and hence different frequencies) will have a different associated $\delta \cgw(f)$ (and hence travel times). This is what leads to the stretching/squeezing of the observed signal, where importantly the distance to the source affects how pronounced this effect is. We note that this squeezing and stretching affect is the same as the frequency-domain dephasing introduced in~\cite{Mirshekari:2011yq}, albeit expressed in a different formalism.
In the case of a large distortion in the observed waveform, this can lead to effects analogous to the `inverse chirping' of scalar waves in scalar-tensor theories \cite{Sperhake:2017itk,Aurrekoetxea:2022ika}.
For distances of 100Mpc a value of only $\delta \cgw(f) = 1\times10^{-16}$ is required to add a shift of 1s to a signal.
While 1 second may seem like a small amount, shifting a signal by 1 second in the tens of seconds it takes a signal to move through the LVK observing band is a significant shift, as we will explore in more detail in later sections.

\subsection{Arrival times}
\label{ssec:arrival_times}
A second case of interest is when there is no detectable frequency-dependence in the LVK or LISA bands themselves, yet there is a sharp transition in between the two bands. In that case systems that are first observable in LISA and then enter the LVK band are an excellent probe of the transition. The primary signature here is not in the form of the signal itself, but instead even a small $\delta\cgw$ will result in an `arrival time' of the signal in the LVK band significantly different from what would be predicted from LISA observations of the signal itself.
The idea of multi-band observations---using information
from both LISA and the LVK band to measure the properties of compact binary mergers---was first demonstrated in~\cite{Sesana:2016ljz}. In that work it was predicted that LISA observations could measure the
arrival time of the signal in the LVK band to an
accuracy of tens of seconds. This finding was
later supported by more detailed followup work~\cite{Marsat:2020rtl,Toubiana:2020cqv}.
However, in an independent study~\cite{Klein:2022rbf}, a more pessismistic prediction was provided. There it is found that for stellar-mass black systems, like GW150914, one can predict the arrival time in the LVK band with a precision of $\sim$ hours from observations in the LISA band. 
This discrepancy has not yet been resolved in the
literature\footnote{We thank the authors of \cite{Toubiana:2020cqv, Baker:2022eiz} for related discussions.}. Nevertheless, even an uncertainty on the arrival time of the signal in the LVK band constrained
to hours can in principle be used to place
very strong constraints on $\delta\cgw$ in the case
of multi-band observations, as we explore in section~\ref{sec-probe-joint}.

\subsection{Intrinsic source effects}
\label{ssec:intrins_effects}

We will explore the effect of the waveform stretching/squeezing and arrival time further in the following sections, but first we will briefly state that the emission of gravitational waves is affected when a non-trivial $\cgw$ is present even when one is very near to the source. 
The full derivation can be found in appendix~\ref{app-intrins-eff}, but we show that the intrinsic evolution of the source, in terms of the angular frequency ($\omega$), is changed in the presence of a non-constant $\cgw$ according to
\begin{align}
\dot\omega(t) =  \left(\frac{G_{\rm GW}c}{G_{N} \cgw}\right) \frac{3456^{1/3}}{5} \left(\frac{G_N M_c }{c^3}\right)^{5/3} \omega^{11/3}.
\end{align}
Here $G_{\rm GW}$ is the effective gravitational `constant' $G$ seen by tensor perturbations in the metric and $G_{N}$ is the `standard' gravitational constant as e.g. entering the Poisson equation or Kepler's laws\footnote{In other words, $G_N$ is the gravitational constant relevant for the interaction between (test) masses, so e.g. for the computation of planetary orbits or torsion balance tests, while (in the context of the Horndeski theories considered here) gravitational waves feel a different effective gravitational constant $G_{\rm GW}$ due to the presence of the extra scalar degree of freedom. For Horndeski theories these satisfy the relationship $G_N/G_{\rm GW} = \cgw^2/c^2$~\cite{Jimenez:2015bwa} -- see appendix \ref{app-intrins-eff} for details.}.
This is the standard general relativity solution, computed to leading order (``0 Post-Newtonian order''), with an additional term containing $\cgw$ and $G_{\rm GW}$. However, as shown in appendix~\ref{app-intrins-eff}, this term/correction will be  negligible in comparison to the arrival time delays (in the region we are exploring), so we will neglect this effect in what follows.

\section{Probing \texorpdfstring{$\cgw$}{[cgw]} in the LVK band} \label{sec-probe-LIGO}

The joint observation of GW170817 and GRB 170817A allowed for a simple, but very powerful, test of the speed of gravity. The peak of gravitational-wave emission
and the gamma-ray burst arrived at the Earth within seconds of each other. If one assumes that these were emitted from the source at roughly the same time, then any deviation between their arrival times at the Earth would be due to differences in the speed of the two messengers. In the
absence of any observable delay, a very strong bound on $\cgw$ can be placed. Indeed, this observation constrained $\delta \cgw$ in the 10 - 1000Hz frequency band to $\lesssim 3\times 10^{-15}$~\cite{LIGOScientific:2017zic}.

This limit was placed assuming a \emph{constant} $\cgw$.
In the scalar-tensor theories we consider here we assume that
$\delta \cgw$ is allowed to be nonzero, but it must be asymptoting to 0 in the $> 10\mathrm{Hz}$ frequency band in which GW170817 was observed by LIGO and Virgo.
We will demonstrate that a much tighter constraint on $\delta \cgw$ across the LVK sensitive band can be placed whenever there is a frequency-dependent $\cgw$.
To do this we will first formulate how we will model the gravitational-wave signal that ground-based observatories would see assuming a variable $\cgw$,
we will then define how we will model our specific example, and then use these models to identify what range of $\cgw$ is consistent with the data observed around GW170817.

\subsection{Including waveform stretching/squeezing in the gravitational wave signal}

We first need to define how we will apply a frequency dependent $\delta \cgw$ to the waveforms that we will be using. As discussed in section~\ref{ssec:intrins_effects}, we neglect any intrinsic source
evolution effects.
However, the variable travel time delay as the waveform evolves from 10Hz towards 1000Hz and merger is not negligible, and we must incorporate this effect into our waveform models.

To do this we begin by using the ``TaylorF2" post-Newtonian waveform \cite{Buonanno:2009zt} as implemented in \texttt{lalsuite}~\cite{lalsuite}, and apply a frequency-dependent time shift to it. We first identify time before merger ($\tau$)
as a function of frequency using the leading order general relativity approximation
\begin{align}
f = \left(\frac{256}{5 c^3} M_c^{5/3} \pi^{8/3} \tau\right)^{-3/8}.
\end{align}
We then identify the time delay as a function of frequency according to
\begin{align}
\delta \tau(f) =  \frac{{D_L}}{c} \delta \cgw(f).
\end{align}
From these two expressions we can get $\delta \tau(\tau)$---the arrival time delay as a function of time before merger---which
can be used to produce the timeshifted waveform we might expect to observe under these assumptions. We apply the frequency dependent time shift
by Fourier transforming
the frequency-domain waveform into the time-domain, applying the time shift to all sample points, and interpolating the waveform back to an equally
spaced timeseries for use in analysis.

We note that our interpolation technique is computationally expensive.
The technique of adding this squeezing and stretching in terms of
phase offset in the frequency domain~\cite{Mirshekari:2011yq} is more
efficient, but would need to be generalized from the power-law case
considered there (although we do use a power-law approximation in this section, we use the form of equation~\ref{GW_speed} in later sections).
We also only consider the dominant mode of gravitational-wave emission in our studies, we refer the reader to \cite{Ezquiaga:2022nak} for a discussion on how higher-order modes could be included.

\subsection{Parameterizing \texorpdfstring{$\delta \cgw(f)$}{[delta cgw]}}

We next need a functional form of $\delta \cgw(f)$. We use the formalism
we discussed in section~\ref{subsec-template} in the limit where $\delta \cgw(f)$ is asymptoting to 0, i.e. in what we previously called the $k \gg k_\star$ limit. We model this
asymptotic behaviour with two parameters---$\delta \cgw(f_{\mathrm{ref}}) $ and $\sigma$---according to
\begin{align}
\delta \cgw(f) = \delta \cgw(f_{\mathrm{ref}}) \times \frac{f_{\mathrm{ref}}^{2\sigma}}{f^{2\sigma}}.
\end{align}
Here e.g. $\delta \cgw(30)$ corresponds to the value of $\delta \cgw$ at $30 \; \Hz$, and we assume that $\delta \cgw$ is approaching zero
over the band of observation following a power law with slope $2\sigma$. $f_{\mathrm{ref}}$ is some reference frequency, which we stress is distinct from $f_\star$.
Note that this emergent power law behaviour 
neatly links up with power-law parametrisations for (modifications of) the dispersion relation previously employed in the literature \cite{Mirshekari:2011yq,LIGOScientific:2018dkp,Baker:2022rhh}. However, it is worth emphasizing that in our context this behaviour only emerges asymptotically and it will eventually be important to keep track of the full functional form of \eqref{GW_speed} when inferring constraints on $\delta\cgw(f)$ throughout the entire range of frequencies considered here. We will do so in the following sections.

\subsection{Approximate measurement of observable \texorpdfstring{$\cgw$}{[cgw]} deviation}

\begin{figure}[t!]
\begin{center}
\includegraphics[width=0.99\linewidth]{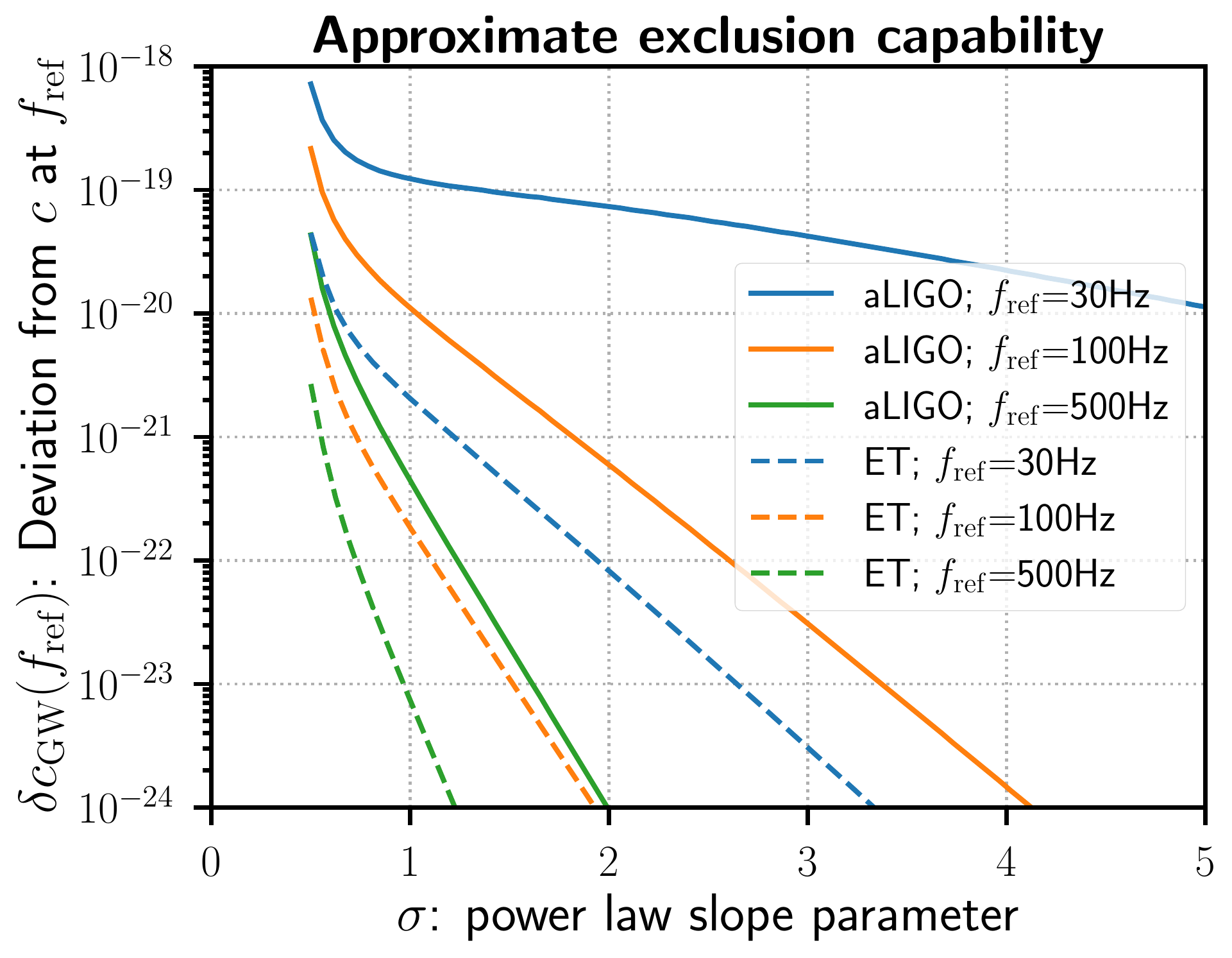}\\
\vspace{0.2cm}
\includegraphics[width=0.99\linewidth]{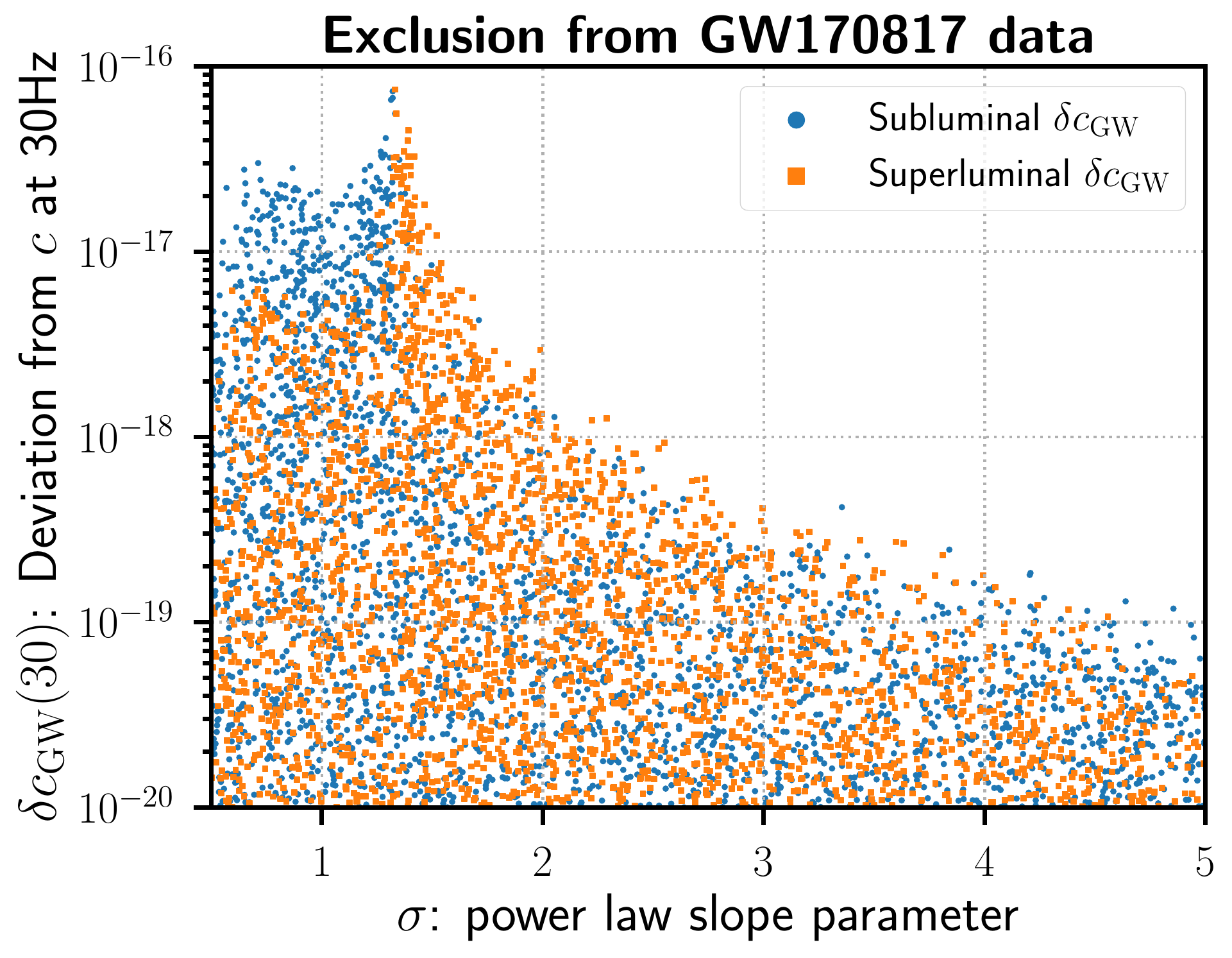}
\caption{\label{fig:gw170817_fisher_exclusion}
(Top)
A prediction of how well we would be able to constrain $\delta \cgw$ at 30Hz, 100Hz and 500Hz as a function of the power law slope parameter $\sigma$. This power law models how quickly $\delta \cgw(f)$
asymptotes to 0 as the frequency increases. This prediction assumes a signal with the same masses (1.36$M_{\odot}$ for both bodies) as GW170817, and the same signal-to-noise ratio
(32.4). The Advanced LIGO noise curve and the Einstein Telescope predicted noise curve~\cite{lalsuite} are used while computing these predictions. (Bottom) Bayesian inference exclusion posterior points, projected down to $\sigma$ and $\delta \cgw(30)$.  An animated version of this panel can be found at 
\href{https://icg-gravwaves.github.io/probing_speed_of_gravity/Figure_2B/figure2b.mp4}{THIS LINK}
}
\end{center}
\end{figure}

We then want to quantify if such a deviation in the waveform is observable with respect to a waveform where always $\cgw = c$.
A way to approximate what values of $\delta \cgw(f_{\mathrm{ref}})$ and $\sigma$ are required to produce a distinguishable effect is to use the "distinguishability criteria" from~\cite{Lindblom:2008cm}.
This states that two waveforms are considered to be distinguishable if, after subtracting one from the other, the residual has a signal-to-noise ratio that is larger than one. That is if
\begin{equation}
\left\langle h_2 - h_1 | h_2 - h_1 \right\rangle \geq 1
\end{equation}
then two waveforms could be distinguished from each other. This uses the definition of the noise-weighted inner product between two waveforms $h_1$ and $h_2$,
which is commonly used in gravitational-wave data analysis
\begin{equation}
\left\langle h_1 | h_2 \right\rangle = 4 \Re \int_{0}^{\infty} \frac{h^{*}_1(f) h_2(f)}{S_n(f)} df.
\end{equation}
Then, to find the level of deviation which is observable, we first choose a fiducial physical system and generate the waveform assuming GR.  We then generate the same waveform, but assuming some parameterized deviation from GR and measure the "distinguishability criterion". We then vary the values of the GR-deviation parameters until we get a value of 1. We note that this measure can provide overly optimistic predictions in the case where there are correlations between parameters, for example it might be possible to change other physical parameters to mimic the waveform stretching/squeezing.
We will check the validity of this approximation in the following
subsection.

In the top panel of Figure~\ref{fig:gw170817_fisher_exclusion}, we demonstrate this for a signal with the same masses and signal-to-noise ratio as GW170817, and using the Advanced LIGO noise curve.
This allows us to predict that such a signal would allow us to constrain $\delta \cgw$ at 30Hz to be no larger than $\sim 10^{-19}$, for a slope with power law $\sigma=1$. For larger values
of $n$ the exclusion quickly becomes much tighter. We also predict
the exclusion of $\delta \cgw$ at 100Hz and 500Hz. Exclusions at
these frequencies are considerably tighter, as would be expected
when we model $\delta \cgw$ as asymptoting to 1.
Finally, we predict the exclusion that would be possible if GW170817
were observed with the future Einstein Telescope observatory, these
exclusions are around two orders of magnitude tighter at $\sigma=1$
and decay considerably quicker, reflecting the improved sensitivity
of Einstein Telescope at lower frequencies. This figure is generated
assuming a subluminal $\cgw$, however we find that the results are
identical to within numerical precision when using a superluminal
$\cgw$\footnote{In the next section we do find slightly different behaviour between subluminal and superluminal $\cgw$. This is because a partial 
degeneracy between source masses and the deviation parameters becomes important, and at some points we see divergence between the two cases.}.

For full reproducibility of this, and other figures in this work,
the code used to produce this figure can be found by \href{https://icg-gravwaves.github.io/probing_speed_of_gravity/}{clicking on this link}.

\subsection{Bayesian parameter estimation constraints of \texorpdfstring{$\cgw$}{[cgw]} using GW170817}

We next consider what constraints can be placed using the recorded strain data
around the GW170817 observation when a frequency-varying $\cgw$ is assumed.
This will allow us to verify the predicted curve in Figure~\ref{fig:gw170817_fisher_exclusion} and check to see if there are
correlations between the parameters parameterizing $\delta \cgw$ and other
physical parameters of the system.
We use Bayesian inference to place bounds on the $\sigma$ and $\delta \cgw(30)$ parameters. We take the
publicly released data from GWOSC from around GW170817~\cite{LIGOScientific:2019lzm} and follow closely the analysis described in~\cite{Nitz:2020oeq}.
We take the known sky location from the optical counterpart
to GW170817~\cite{LIGOScientific:2017ync} and assume a luminosity distance of 40Mpc~\cite{Cantiello:2018ffy} when applying the frequency-dependent time shift. We do not consider spins in the
analysis as neutron star spins are expected to be small~\cite{Brown:2012qf}, which allows
us to reduce computational cost and we also do not include tidal terms
in the waveform.

We choose a prior on $\sigma$ that is uniformly distributed between 0.5 and 5, and a uniform in log prior on $\delta \cgw(30)$ between $1\times10^{-13}$ and $1\times10^{-20}$. Other prior ranges match those of~\cite{Nitz:2020oeq}. We use
the emcee parallel tempered sampler~\cite{Foreman-Mackey:2012any} within PyCBC inference~\cite{Biwer:2018osg} to generate results.
We run the analysis twice, once allowing for only subluminal values
of $\cgw$ and once allowing for only superluminal values of
$\cgw$. The results are then combined for visualization. For full
reproducability, the configuration files and code versions  \href{https://icg-gravwaves.github.io/ian_harry/probing_speed_of_gravity/}{can be found in our data release page}.

We can see the results of the Bayesian analysis in the bottom panel of Figure~\ref{fig:gw170817_fisher_exclusion}. We observe
that the exclusion plots strongly depend on the
frequency dependence of $\cgw$, i.e., on the power
law slope $\sigma$ in Equation~\ref{sTemp_powerlaw}. The vast
majority of samples we obtain have a $\delta \cgw$ at 30Hz smaller than $1\times10^{-16.5}$, although there is one curious spike where $\sigma \sim 1.5$, where a shift in the masses of the source, coupled with this particular frequency-dependent time shift, seems to reproduce the GW170817 waveform well. This occurs for both
superluminal and subluminal $\cgw$, although the shifted source
masses differ in the two cases. 
We also provide an animated version of this plot, where we show how
the constraints improve as a function of frequency. We find that
all samples have $\delta \cgw$ smaller than $1\times10^{-17}$ at 100Hz, and all samples are smaller than $1\times10^{-18}$ at 500Hz.
We can also compare the results between the two panels of
Figure~\ref{fig:gw170817_fisher_exclusion} to validate the accuracy
of the distinguishability criterion used in the top panel. While we
do notice that the top panel is more optimistic than the bottom panel,
it does agree to within 1-2 orders of magnitude over the range of
parameters considered. We therefore consider the distinguishability
criterion sufficient for making qualititative predictions and we will
use this again in future sections.

These constraints are considerably stronger than the limit of $\lesssim 3 \times 10^{-15}$ placed with GW170817~\cite{LIGOScientific:2017zic}. This can be understood by remembering that this limit was based off of a time delay of $\sim 5$ seconds.
However, a shift of even milliseconds in a gravitational-wave signal can be observable with matched-filtering and so we are able to probe
much smaller values of $\delta \cgw$.

The LVK collaboration papers have also considered a variable
gravitational-wave speed in terms of a ``modified dispersion relation''~\cite{LIGOScientific:2017bnn, LIGOScientific:2018dkp,LIGOScientific:2019fpa,LIGOScientific:2020tif, LIGOScientific:2021sio}. Bounds in these papers have not been directly quoted
in terms of $\delta \cgw$. However, it is possible to translate between the constraining power on, for example, the graviton mass quoted in these papers and infer the implied constraining power for $\delta \cgw$\footnote{We thank Nathan Johnson-McDaniel for pointing this out to us and for discussions about how the LVK results compare to our Figure~\ref{fig:gw170817_fisher_exclusion}.}.
The LVK analyses only compute bounds at specific values of $\sigma$, and do not consider $\sigma$ values larger than 1. However, the results from analysing
GW170817 data are compatible with our results, in regions where our analyses overlap~\cite{LIGOScientific:2018dkp}.
Tighter bounds on the graviton mass have been placed on the LVK catalog
using 43 confident binary black hole mergers in~\cite{LIGOScientific:2021sio}. If this data were used to place bounds on $\cgw$ it would result in bounds that are around two orders of magnitude tighter than those quoted here. This further backs up the main point of this section, which is that ground based observations of gravitational-wave mergers can place, and indeed have placed, very tight constraints on $\cgw$ in the $\sim 10 - 1000$Hz region.

\section{Probing \texorpdfstring{$\cgw$}{[cgw]} in the LISA band} \label{sec-probe-LISA}

\begin{figure*}[t]
\centering
\includegraphics[width=0.99\linewidth]{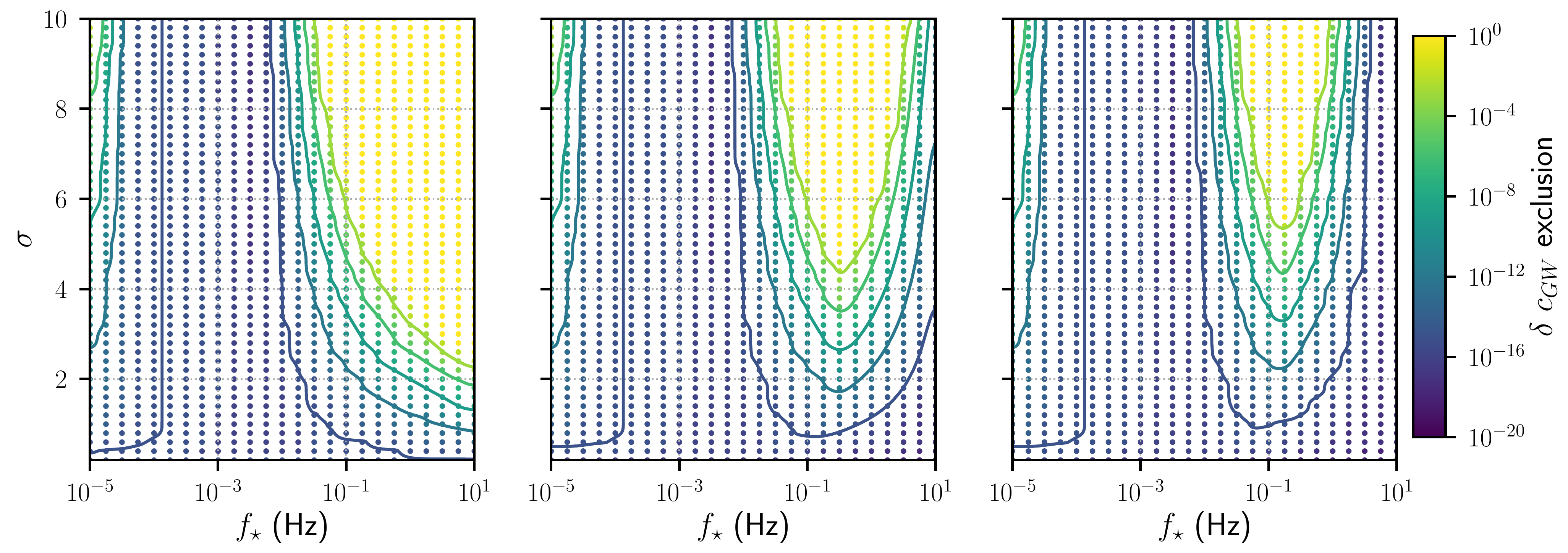}
\caption{\label{fig:exclusion_fisher}
The values of $\delta \cgw^{(0)}$ that would be distinguishable from a purely GR signal, as a function of the transition frequency ($f_{\star}$) and the steepness of the transition ($\alpha$). (Left) This is computed assuming a single LISA observation of a system of two supermassive black holes each having a mass of $4.154\times 10^6 M_{\odot}$ at a distance of 1Gpc. (Middle) Computed assuming the combination of an LVK observation of a GW170817-like source and a single LISA observation of two supermassive black holes. (Right) Computed assuming the combination of an ET observation of a GW170817-like source and a single LISA observation of two supermassive black holes. In all cases contour lines correspond to $\delta \cgw^{(0)} = 10^{-3}, 10^{-6}, 10^{-9}, 10^{-12}$ and $10^{-15}$. 
}
\end{figure*}

In the previous section we have demonstrated that we can place much tighter constraints upon deviations from general relativity
in the LVK band if we assume a model where the speed of gravity diverges from the speed of light, but varies as a function of frequency. Even a very small variation
as a function of frequency can cause a measurable frequency-dependent delay in arrival time in the observed gravitational-wave signal.
Similar tests would also be possible for observations in the 1mHz - 100mHz with future space-based observatories, such as LISA~\cite{LISA:2017pwj} and TianQin~\cite{TianQin:2015yph}.
Here we focus on the LISA observatory. Perhaps the most useful astrophysical source of gravitational waves for probing deviations from general relativity with LISA 
is the merger of two supermassive black holes. Such systems would have very large signal-to-noise ratio, would have relatively small
mass ratios---where reliability of waveform models is highest---and would cover a broad range of frequencies in the scale of O(few years) that LISA might observe
such a source~\cite{Klein:2015hvg, Katz:2018dgn}.

Within the LISA context, bounds on $\delta\cgw$ have previously been considered in two different settings. First, \cite{Littenberg:2019mob} investigated testing a frequency-independent $\delta\cgw$ in the LISA band with a multi-messenger observation akin to GW170817. They forecasted a resulting bound $|\delta\cgw| \lesssim 10^{-12}$ in the event of a non-detection. As before, here we instead focus on candidate signatures of a frequency-dependent $\delta\cgw$ and (given our findings in the LVK band above) expect to place significantly stronger bounds on such a signal. Secondly, while this paper was being completed \cite{Baker:2022rhh} also investigated a frequency-dependent $\cgw$ in the LISA band, similarly motivated by the observation of \cite{deRham:2018red} that a frequency-dependent $\cgw$ can lead to interesting phenomenology close to the LISA band and exploring its effect on waveform models and on (forecasted) parameter constraints. 
In terms of $\delta\cgw$ constraints, their results are driven by redshift-induced frequency dependence (e.g. that the frequency of a monochromatic wave differs at the source and at the observer in an expanding Universe). 
This effect also is a generic consequence of a frequency-dependent $\cgw$ and \cite{Baker:2022rhh} uses this to place bounds on $|\delta\cgw| \lesssim 10^{-4}$. Here we show that these bounds can be improved by several orders of magnitude, more specifically to $|\delta\cgw| \lesssim 10^{-17}$, by considering the effects of $\cgw \neq c$ outlined in section \ref{sec-signal}. 

To demonstrate the potential capability of LISA to observe such effects we again use the distinguishability criterion introduced in the previous section.
Here we consider a system with a pair of black holes both having a mass of $4.154\times 10^6 M_{\odot}$
(the mass of the black hole at the centre of the Milky Way). We situate the system at a distance of 1 Gpc, which we take to be at a redshift of 0.20.
We assume that the sky position
and orientation of the system is such that it will induce the optimal response for the LISA detector and we assume that we observe the system from an initial gravitational wave
frequency of $5\times 10^{-5}$Hz, corresponding to roughly 1.5 months before merger, the period in which the majority of the signal power would be accrued.
We model the LISA sensitivity curve using the \texttt{LISA Sensitivity} code\footnote{\url{https://github.com/eXtremeGravityInstitute/LISA_Sensitivity}}.
Such a system would be recovered with a signal-to-noise ratio of around 5000 with LISA. It is possible that the transition frequency would occur
within the LISA band of observation, so rather than modelling $\delta \cgw$ as an asymptoting function, we use the 3-parameter model defined earlier in equation~\ref{GW_speed}.
We then, as a function of the transition frequency ($f_{\star}$) and the steepness of the transition ($\alpha$), predict the values of $\delta \cgw^{(0)}$ that would be distinguishable
from a signal modelled assuming GR.

The result of this can be seen in Figure~\ref{fig:exclusion_fisher}. We can see that in the absence of a detection of a deviation from GR with such an observation, it would be
possible to place very tight bounds over a large range of the parameter space that we consider. However, if $\cgw$ has a steep transition with a transition frequency above
$0.01$ to $0.1$Hz it would produce a signal in LISA that looks almost identical to a purely GR signal. 
It is important to remember that the detectability criterion we apply
here was demonstrated to be optimistic by 1 to 2 orders of magnitude
in the LVK band, and we might expect a similar behaviour here as well. Nevertheless, this will not affect the main features of our results
shown in Figure~\ref{fig:exclusion_fisher}.
In this context also note that the bounds forecasted here are for a single source and, from the analogous LVK band discussion in the previous section, one may expect these bounds to improve in a similar manner when considering a catalog of LISA observations.

\section{Probing \texorpdfstring{$\cgw$}{[cgw]} with joint LVK/LISA observations} \label{sec-probe-joint}

It is interesting to compare how the bounds placed from simulated observations in the LVK band would combine with analogous bounds in the LISA band.
If we re-compute our approximate bounds on $\cgw$ used in Figure~\ref{fig:gw170817_fisher_exclusion} in terms of the 3 parameters used in the previous section, it is easy to combine these plots and produce a visualization of the area of parameter space that would be excluded from observations in both bands. This can be seen in Figure~\ref{fig:exclusion_fisher}. We can see that this combined
exclusion includes much tighter constraints at larger values of the transition frequency, and would place very strong constraints for $\sigma < 2$ over the full range of transition frequencies considered. However, it would still permit cases where the transition frequency is between the two bounds (around 0.1 to 1 Hz) with a very steep transition curve. For completeness, we also show in
Figure~\ref{fig:exclusion_fisher} how the constraints would improve
with the observation of a GW170817-like source with the Einstein
Telescope. Finally, we also include in Appendix~\ref{app-figure-a2}
the uncombined exclusions, with only an LVK or ET observation of a 
GW170817-like source. Indeed, there we see that LVK band observations by themselves already strongly constrain $\delta\cgw$ throughout the LISA band for $\sigma = 1$.

In addition to observations made with only ground-based, or only with space-based observatories, it will be possible, in the coming decades, to observe a gravitational-wave source
\emph{both} with ground-based and space-based observatories. This was first explored in~\cite{Sesana:2016ljz} where it was demonstrated that a GW150914-like merger would
be observable to LISA in the years before it merged in the ground-based observation window. Such \emph{multi-band} observations offer the ability to perform particularly stringent
tests of general relativity from observations of sources over a wide range of frequencies. In particular, consider the case that $\cgw$ asymptotes to 1 before entering the
ground-based observation window, and $\cgw$ asymptotes to a constant $\cgw(0) \neq c$ in the LISA band of observation (at levels of accuracy sufficient to evade the detection of any frequency-dependence in the LVK and LISA bands themselves -- see previous sections).  Such a deviation would
not be directly observable in either band, but in the gap between the signal leaving the LISA band and being observed in the LVK band there would be a shift in the travel time required for the signal to reach an observer.
Relating this to our model fit of equation~\ref{GW_speed} this would require a large value of $\sigma$ with 
$f_\star$ in the frequency gap between the LISA and LVK bands, which
is the region that is poorly constrained in Figure~\ref{fig:exclusion_fisher}.

If we consider GW150914 at a distance of 400Mpc, the travel time, in the Earth's reference frame, for the signal to reach the Earth is $4 \times 10^{16}$ seconds. Then, as with the previous cases, we can see that even very tiny changes in $\cgw$ will produce large variations in the arrival time. To produce time shifts of $\sim 10000$ seconds---roughly the accuracy at which an observation in the LISA band would be able to constrain the merger time as predicted in~\cite{Klein:2022rbf}\footnote{We remind the reader that there are inconsistent predictions of the ability to measure the arrival time of such a signal in the LVK band with LISA, as we discussed in section~\ref{ssec:arrival_times}}---would require $|\delta\cgw| = 3 \times 10^{-13}$. To produce time shifts of $\sim 10$ seconds---as predicted in~\cite{Sesana:2016ljz,Marsat:2020rtl,Toubiana:2020cqv}---would require $|\delta\cgw| = 3 \times 10^{-16}$. More generally speaking, if in the future the merger time in the LVK band can be predicted with an accuracy of $\sim 10^x$ seconds from LISA band observations, $\delta \cgw \sim 10^{-14+x-y}$ will be discernible for a source at $10^y$ Mpc.

An interesting observation can be made when considering larger values of $|\delta\cgw|$. In particular the time delay due to $|\delta\cgw|$ changing between the LISA and LVK bands can quickly become very large. A value of only $|\delta\cgw| = 8 \times 10^{-10}$ is required before the time delay is larger than a year. In such cases, with a subluminal $\cgw$, we might observe the signal with LISA at \emph{the same time} as the signal arriving at ground based observatories. For values of $|\delta\cgw|$ that are much larger than this, we would not observe a multi-band signal at all, and indeed the absence of multi-band signals, when such observations are expected, could be an indicator of $|\delta\cgw|$ being large in the LISA band.
With existing upper bounds on $\cgw$ relevant for the LISA band imposing $|\delta\cgw| \lesssim 10^{-2}$ \cite{Jimenez:2015bwa}, 
this means there is in principle a window $|\delta\cgw| \sim 10^{-8}-10^{-2}$ for which, given a sufficiently fast transition in between the LVK and LISA bands, the absence of multi-band signals would be the main signature in near-future measurements.
This would also be an especially interesting target for proposed future experiments targeting this intermediate region, such as AEDGE \cite{AEDGE:2019nxb}.
Finally note that, while this draft was in internal review, we were made aware of ongoing work \cite{Baker:2022eiz} that will also explore similar multiband constraints.

\section{Conclusions} \label{sec-conc}

\ni {\bf Summary}: In this paper we have explored how frequency-dependent transitions in the speed of gravitational waves $\cgw$ -- a generic consequence in large classes of dark energy theories -- can be constrained by current and future observations in the LIGO/Virgo/KAGRA (LVK) band, in the LISA band and with joint observations in both bands. Since such dark energy-related transitions naturally occur close to those two bands, the observations place powerful constraints on $\cgw$. We would like to highlight the following key results:
\begin{itemize}
\item We find that deviations away from $\cgw = c$ can be constrained down to a level of $|\delta\cgw| \sim 10^{-17}$ in {\it both} the LVK {\it and} LISA bands even for mild frequency-dependence, much stronger than existing bounds for frequency-independent $\cgw$ from GW170817 or indeed analogous forecasted bounds for multi-messenger observations in the LISA band \cite{Littenberg:2019mob}. Our constraints are driven by $\cgw$-induced frequency-dependent time shifts in the observed signals and we have discussed in detail how they depend on the precise frequency(-ies) associated with the transition as well as on its functional form. Fig. \ref{fig:exclusion_fisher} summarises the main constraints we find.
\item We have identified a class of interesting very fast transitions taking place around frequencies of $\sim 10^{-2} - 10 \; \Hz$, which can almost completely evade the aforementioned bounds when they proceed sufficiently quickly\footnote{In terms of a power law scaling, this e.g. means $\cgw = c$ is approached at large frequencies as $f^{-2\sigma}$, where $\sigma \gtrsim 5$.}. However, joint observations of sources visible in both LVK and LISA bands would be able to either: 1) constrain $|\delta\cgw| \lesssim 10^{-15}$, when LVK observations `see' sources at the times expected from prior LISA observations of the same sources within expected uncertainties (note that this constraint can weaken to $|\delta\cgw| \lesssim 10^{-12}$ for more pessimistic forecasted uncertainties -- see section \ref{sec-probe-joint} for details), 2) measure $|\delta\cgw|$ anywhere in the range $|\delta\cgw| \sim 10^{-15}-10^{-9}$, or 3) indicate that $|\delta\cgw| \sim 10^{-8}-10^{-2}$, when LVK observations do not `see' any expected counterparts for sources expected from prior LISA observations within ${\cal O}$(years).
\item As a precursor for deriving these constraints, we have constructed a number of theoretically well-motivated templates for a fiducial frequency-dependent $\cgw$. Underlying most of our results is a simple three-parameter ansatz that captures the essential features of candidate transitions (see section \ref{sec-freq}), but we collect a discussion of how to construct more sophisticated templates for future analyses in appendix \ref{app-templates}.
\end{itemize}

\ni {\bf Implications for dark energy}:
In the immediate aftermath of GW170817 the consequences of tight bounds on $\cgw$ -- $|\delta\cgw| \lesssim 10^{-15}$ -- on dark energy theories were explored in detail (see e.g. \cite{Baker:2017hug,Creminelli:2017sry,Sakstein:2017xjx,Ezquiaga:2017ekz} and references therein), assuming these bounds can simply be ported to much lower cosmological frequencies. The resulting conclusions would be significantly more robust in the presence of an analogous precision measurement at LISA frequencies, as discussed and forecasted here. This is because we can consistently describe both cosmological and LISA scales in dark energy theories that lead to $\cgw \neq c$ on cosmological scales, while this is much more challenging for LVK scales \cite{deRham:2018red}. 

In the context of Horndeski gravity explored in this paper, a tight bound on $\delta\cgw$ from LISA would therefore firmly reduce the set of surviving scalar-tensor dark energy theories to \cite{Baker:2017hug,Creminelli:2017sry,Sakstein:2017xjx,Ezquiaga:2017ekz} (also see \cite{Amendola:2012ky,Amendola:2014wma,Deffayet:2010qz,Linder:2014fna,Raveri:2014eea,Saltas:2014dha,Lombriser:2016yzn,Lombriser:2015sxa,Jimenez:2015bwa,Bettoni:2016mij,Bettoni:2016mij,Sawicki:2016klv} for closely related prior work)
\begin{align} \label{survivor-lag}
{\cal L} = G_2(\phi, X) + G_3(\phi,X)\Box \phi + G_4(\phi)R,
\end{align}
where the $G_i$ are free functions of the scalar $\phi$ and its first derivative via $X \equiv -\tfrac{1}{2}\nabla_\mu \phi \nabla^\mu \phi$. Conversely, within the Horndeski context a detection of a non-zero $\delta\cgw$ would constitute a measurement of significant $G_4$ and/or $G_5$ interactions contributing to $\cgw$. 
Note that, by constraining the templates discussed here and in the event of a detection of non-zero $\delta\cgw$, not only would we obtain information about $\delta\cgw$ itself and hence the value of $\cgw$ on cosmological scales, but also about the $f_\star$ and $\sigma$ parameters detailed in section \ref{sec-freq}. $f_\star$ encodes information about the energy/frequency-scale associated with a UV completion for dark energy, while $\sigma$ encodes information about the specific novel physics and interactions entailed by such a candidate UV completion. 
\\

\ni {\bf Implications for future gravitational wave searches}:
Throughout most of this paper our search for and constraints on $\cgw$ with separate or combined, present and or forecasted LVK/LISA observations has been primarily motivated by large classes of dark energy theories that generically lead to a frequency-dependent $\cgw$ in or close to those bands. What do our results imply beyond this dark energy-specific context? Ultimately a non-zero $\delta\cgw$ would be evidence for novel gravitational interactions, new physics, that provide a non-trivial medium for gravitational waves to travel through. The template(s) we have worked with here do assume 1) an asymptotically constant $\cgw$ at very low frequencies and 2) a frequency-dependent transition to $\cgw = c$ at some frequency $f_\star$, whose form we have motivated by general field-theoretic arguments. Importantly, this means no further dark energy-specific input affects the templates themselves, so these can be implemented in future searches and offer a robust probe of any novel physics affecting $\cgw$ in the LVK/LISA bands and (at least approximately) consistent with the above two assumptions, whether dark energy-related or otherwise. 
%\\

\section*{Acknowledgments}
\vspace{-0.1in}
\noindent 
We especially thank Nathan Johnson-McDaniel and Scott Melville for several helpful discussions and shared insights. We also thank Tessa Baker, Anson Chen, Claudia de Rham, Macarena Lagos, Eugene Lim, Mauro Pieroni, Anand Sengupta and Gianmassimo Tasinato for useful discussions and comments on a draft.
JN is supported by an STFC Ernest Rutherford Fellowship (ST/S004572/1). IH is supported by STFC grants ST/T000333/1 and ST/V005715/1.
In deriving the results of this paper, we have used: xAct~\cite{xAct}, PyCBC~\cite{alex_nitz_2022_6583784},
LALSuite~\cite{lalsuite},
emcee~\cite{Foreman-Mackey:2012any},
numpy~\cite{Harris:2020xlr} and scipy~\cite{Virtanen:2019joe}.

\section*{Data availability}
\vspace{-0.1in}
\noindent 
All data used in this work, and the information
necessary to fully reproduce any results and figures
presented here, can be found at our data release
page {\url{https://icg-gravwaves.github.io/probing_speed_of_gravity/}}.

\section*{Open access}
\vspace{-0.1in}
\noindent 
This article is licensed under a Creative Commons Attribution 4.0 International License, which permits use, sharing, adaptation, distribution and reproduction in any medium or format, as long as you give appropriate credit to the original author(s) and the source, provide a link to the Creative Commons licence, and indicate if changes were made. The images or other third party material in this article are included in the article’s Creative Commons licence, unless indicated otherwise in a credit line to the material. If material is not included in the article’s Creative Commons licence and your intended use is not permitted by statutory regulation or exceeds the permitted use, you will need to obtain permission directly from the copyright holder. To view a copy of this licence, visit {\url{http://creativecommons.org/licenses/by/4.0/}}.

\appendix

\section{\texorpdfstring{$\cgw$}{[cgw]} in Horndeski gravity} \label{app-horn}

In section \ref{sec-freq} of the main text we focused on a specific example theory, namely \eqref{example-lag}. However, the key features relevant for this paper remain true for the much more general class of Horndeski theories, 
the most general Lorentz-invariant scalar-tensor action that gives rise to second order equations of motion. It is described by the following action
\begin{eqnarray}\label{Horndeski_action}
S_H=\int \mathrm{d}^4x \sqrt{-g}\left\{\sum_{i=2}^5{\cal L}_i[\phi,g_{\mu\nu}]\right\},
\end{eqnarray}
where we write the scalar-tensor Lagrangians ${\cal L}_i$ (for a scalar $\phi$ and a massless tensor $g_{\mu\nu}$) as
\begin{align}
{\cal L}_{2} & = \Lambda_2^4 \, G_2~, \quad\quad\quad\quad {\cal L}_{3} = \frac{\Lambda_2^4}{\Lambda_3^3} G_{3}\cdot[\Phi]\,, \nn \\
{\cal L}_{4}  & = \frac{\Lambda_2^8}{\Lambda_3^6} G_{4} R + \frac{\Lambda_2^4}{\Lambda_3^6} ~ G_{4,X} \left( [\Phi]^2-[\Phi^2] \right)\,,   \label{Horndeski_lagrangians} \\
{\cal L}_{5} & = \frac{\Lambda_2^{8}}{\Lambda_3^9} G_{5}G_{\mu\nu}\Phi^{\mu\nu}-\frac{1}{6} \frac{\Lambda_2^4}{\Lambda_3^9} G_{5,X}([\Phi]^3 -3[\Phi][\Phi^2]+2[\Phi^3]). \nn
\end{align}
Here we adopt a dimensionless definition $X \equiv -\tfrac{1}{2}\nabla_\mu \phi \nabla^\mu \phi/\Lambda_2^4$ for what is essentially the scalar kinetic term, and we have defined $\Phi^{\mu}_{\;\; \nu} \equiv  \nabla^\mu \nabla_\nu\phi$. The $G_i$ are dimensionless functions of $\phi/\MPl$ and $X$, and $G_{i,\phi}$ and $G_{i,X}$ denote the partial derivatives of the $G_i$ (with respect to these dimensionless arguments). Square brackets denote the trace, e.g. $ [\Phi^2] \equiv \nabla^\mu \nabla_\nu \phi \nabla^\nu \nabla_\mu\phi$, and we have three mass scales: $\MPl,  \Lambda_2$ and $\Lambda$. In cosmology they are conventionally taken to satisfy $\Lambda_2^2 = \MPl H_0$ and $\Lambda_3^3 = \MPl H_0^2$, which ensures that all interactions can give $\Oo$ contributions to the (cosmological) background evolution. General Relativity is recovered when $G_2 = G_3 = G_5 = 0$ and $G_4 = 1$. Note that the example theory \eqref{example-lag} effectively amounts to a minimal choice of $G_2$ and $G_4$, while $G_5 = g(\phi)$ and $G_3 = 0$. For the general Horndeski theory\eqref{Horndeski_lagrangians}, one can then work out the effect on $\cgw$ and find
\begin{align}
\aT &= 2\frac{X}{M^2}\left[2G_{4,X}-2G_{5,\phi}-\left(\frac{\ddot{\phi}}{H_0^2}-\frac{\dot{\phi}H}{H_0^2}\right)G_{5,X}\right] \,.
\label{cGWcosmo}
\end{align}
Here $M^2/2 = G_4-2XG_{4,X}+XG_{5,\phi}-\tfrac{\dot \phi H}{H_0^2} XG_{5,X}$ is the so-called effective Planck mass (albeit dimensionless in the way written here), $H$ is the Hubble scale that measures the expansion of the Universe and satisfies $H = \dot a/a$ (where a is the scale factor of the Universe), and $H_0$ is the value of the Hubble scale today. Overdots denote time-derivatives with respect to proper time $t$ and we recall that, on  a cosmological background, the scalar $\phi$ is a function of time only, i.e. $\phi = \phi(t)$. From this it is clear that the presence of the scalar can affect $\cgw$ through non-trivial $G_4$ and $G_5$ interactions. Conversely, if $G_5$ vanishes and $G_4$ is at most a function of $\phi$ (but not of its derivatives), then $\aT = 0$ and there is no effect on $\cgw$ and no deviation of this speed from the speed of light.

\section{More \texorpdfstring{$\cgw$}{[cgw]} templates} \label{app-templates}

\begin{figure}[t!]
\centering
\includegraphics[width=\linewidth]{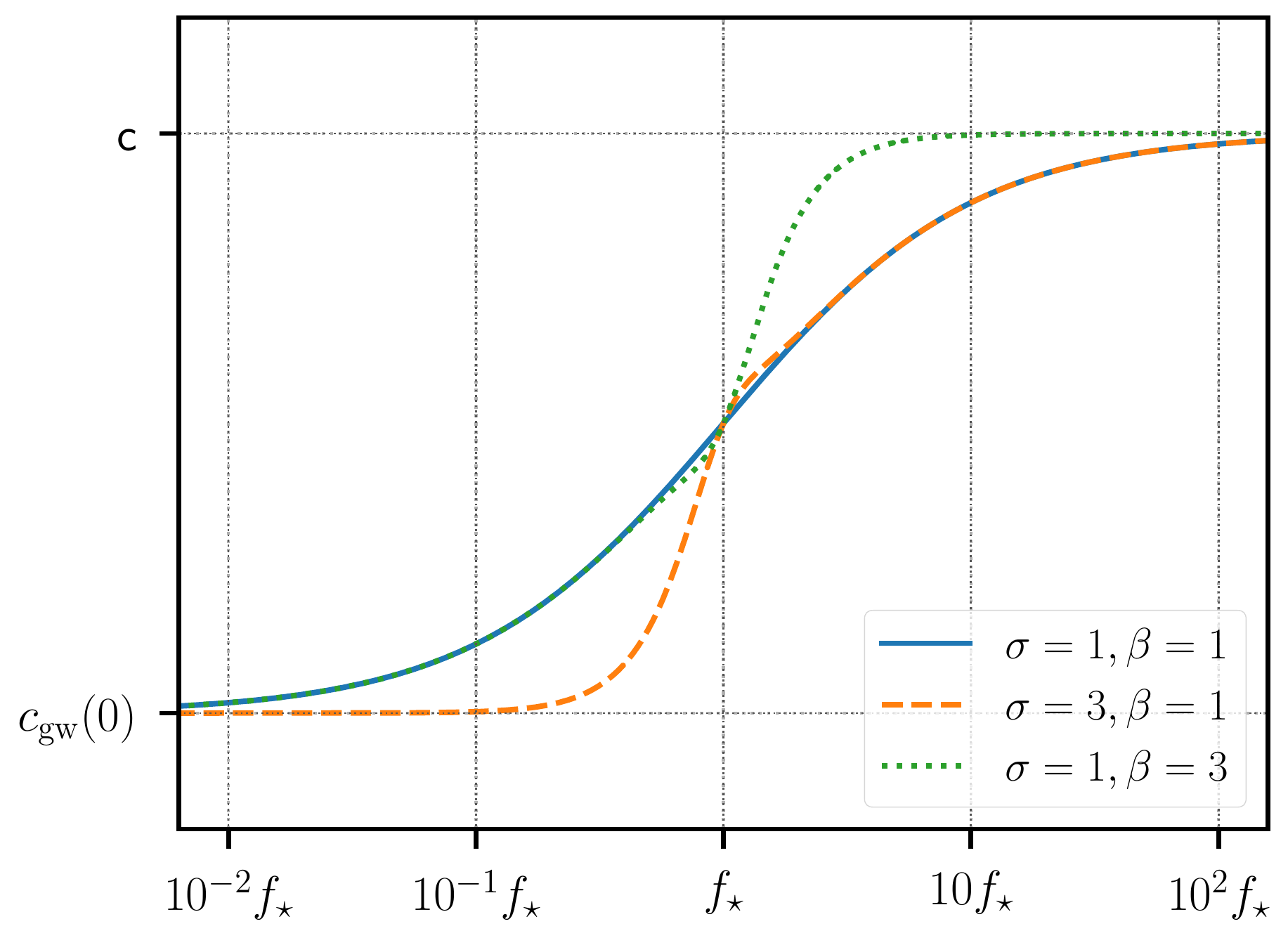}
\caption{
The speed of gravitational waves, $\cgw$, as a function of frequency $f$ for the extended template \eqref{GW_speed_app}. The different curves correspond to different choices of the `steepness' parameters $\sigma$ and $\beta$ that control the asymptotic power law scalings of $\cgw$ -- see \eqref{cgw_power_app}. We have set $\gamma = 6$ for this example. This choice does not affect the asymptotes, but does determine how quickly the transition from a $\sigma$-dominated evolution to a $\beta$-dominated one occurs.}
\label{fig-cgwEvo-app}
\end{figure}

The $\cgw$ template \eqref{GW_speed} we discussed in the main text is controlled by a single steepness parameter $\sigma$ that controls both the small and large $k$ asymptotes. The simplicity of the template results in a symmetric $k^{2\sigma}$ vs. $k^{-2\sigma}$ scaling for those asymptotes, but this can straightforwardly be refined by considering higher order (in the number of parameters) generalisations of this template, in particular ones that allow for different power law scalings in the asymptotic small and large frequency regimes.
Here we will discuss how this can be done more explicitly.

Using the $\Theta_\pm$ functions defined in \eqref{Thetas} we can define
\begin{align}
\hat\Theta(\sigma,\beta,\gamma) \equiv \Theta_+(\gamma)\Theta_+(\beta)+\Theta_-(\gamma)\Theta_+(\sigma),
\end{align}
where it is understood that any $\Theta_\pm$ depends on $f$ and $f_\star$, but we no longer write this explicitly. We can now again build a template for $\cgw$ as before, writing
\begin{align} \label{GW_speed_app}
\cgw(f) = \cgw^{(0)} + (c - \cgw^{(0)})\hat\Theta.
\end{align} 
Note that we require $\gamma > \{ \sigma,\beta\}$, but since it will not affect the large or small frequency asymptotes the precise value of $\gamma$ is immaterial\footnote{Nevertheless note that it does affect the detailed transition evolution around $f_\star$. Since those details are of lesser importance for the phenomenology investigated throughout this paper, we leave a more precise investigation of their effects to future work.}. In this sense \eqref{GW_speed_app} is really a four parameter template controlled by $\{\cgw (0), f_\star, \sigma, \beta\}$. Investigating its asymptotic scalings explicitly, we find
\begin{align}
&k \ll k_\star:  &\cgw - \cgw(0) &\propto k^{2\sigma}, \nn \\
&k \gg k_\star:  &c - \cgw &\propto k^{-2\beta}.
\label{cgw_power_app}
\end{align}
So $\sigma$ and $\beta$ control the relevant power law exponents in the large and small $k$ regime, respectively, allowing different scalings in those two regimes. 
As discussed in the main text, a sufficiently fast (polynomial or otherwise) scaling as $\cgw$ asymptotes back to unity is particularly important for ensuring consistency with measurements in the LVK band. So allowing for this parametric freedom for the large $k$ scaling is especially relevant there, ultimately capturing one specific aspect of the lack of knowledge we have about the precise nature of any would-be UV completion of relevance for our analysis.  

Finally, let us point out that the above template for asymmetric transitions is of course not unique. One example of an alternative template (albeit with less clean asymptotics) is a logistic/Verhulst function of the following type
\begin{align}
\Theta_V(\sigma,\beta) \equiv 1-\left(1+(2^{\beta}-1)\left(\frac{f}{f_\star}\right)^\sigma\right)^{-1/\beta},
\end{align}
from which we can build the template
\begin{align}
\label{GW_speed_app2}
\cgw(f) = \cgw^{(0)} + (c - \cgw^{(0)})\Theta_V.
\end{align} 
Again we have introduced another parameter, $\beta$, that controls the asymmetry in the transition, allowing asymmetric scalings for small and large $k$ in the transition.

\section{Intrinsic source evolution with non-constant \texorpdfstring{$\cgw$}{[cgw]}} \label{app-intrins-eff}

In this appendix we would like to understand how a non-constant $\cgw$ affects GW emission and propagation in the source frame, in particular how the intrinsic evolution of (and power emitted by) a binary system is affected if $\cgw$
deviates from its GR prediction.
We will compute this at leading, 0-PN, order, to demonstrate that these
effects exist, but are negligible in our work. Higher order corrections
would be needed if these effects were important.
Closely following \cite{Jimenez:2015bwa,Maggiore:1900zz}, we can write the quadratic action for the standard metric tensor perturbations $\gamma$ as
\begin{align} \label{tensor_action}
S_h = \frac{1}{64\pi G_{\rm GW}}\int d^4x \left(\dot{\gamma}_{ij}\dot{\gamma}^{ij}-k^2 \cgw^2{\gamma}_{ij}{\gamma}^{ij}\right),
\end{align}
where $G_{\rm GW}$ denotes the effective gravitational `constant' $G$ seen by these tensor perturbations. We can then express the instantaneous power emitted as
\begin{equation} \label{8}
\frac{d E}{d t}= \frac{r^2 c^4}{32 \pi \cgw \Ggw}\int d\Omega \left\langle \partial_t \gamma_{ij} \partial_t \gamma_{ij}\right\rangle\, ,
\end{equation}
where we integrate over solid angle $\Omega$ and average over a region of spacetime much larger than the GW wavelength (denoted by $\langle \dots \rangle$). 
In order to use the above we furthermore solve for the amplitude $\gamma_{ij}$ of the radiated GWs by performing a multipole expansion and working to lowest (quadrupole) order in velocity. Solving the linearised Einstein equation and proceeding along the usual Green's function solution (see e.g. \cite{Maggiore:1900zz}), 
we recover the expression \cite{Jimenez:2015bwa}
\begin{equation}
[\gamma_{ij}]_{quad} \ = \ \frac{2 G_{\rm GW}}{rc^4}  \ddot{Q}_{ij}^{TT} \!\left(t - \frac{r}{\cgw}\right),
\label{solgamma}
\end{equation}
at leading order in velocity. Here we have defined the usual quadrupole moment $Q^{ij} \equiv M^{ij}-\tfrac{1}{3}\delta^{ij}M^k_k$ in terms of the momenta of $T^{00}/c^2$, $M$. Note that, in performing the above computation it is important that the time it takes for a GW to traverse the source is negligible compared to the time variation scale of $\cgw$, which allows us to effectively treat $\cgw$ as constant in the relevant integrals.

Now we would like to work out the {\it total} power emitted, starting from \eqref{8}.
Re-arranging, we obtain
\begin{align} \label{dPdOm}
\frac{dP}{d\Omega} &= \frac{r^2 c^4}{32 \pi \cgw G_{\rm GW}} \left\langle \partial_t \gamma_{ij} \partial_t \gamma_{ij}\right\rangle\ = \frac{G_{\rm GW}}{8\pi \cgw c^4} \left\langle \dddot{Q}_{ij}^{TT} \dddot{Q}_{ij}^{TT}\right\rangle\, .
\end{align}
Here $P \equiv dE/dt$ and we have substituted for $\gamma_{ij}$ using \eqref{solgamma}. Instead of writing this in terms of its transverse-traceless projection $Q_{ij}^{TT}$, we would like to express this in terms of the quadrupole moment of the source $Q_{ij}$ itself. Doing so and integrating over solid angle again, we find the total (quadrupolar) power emitted (at a given time)
\begin{align}
P_{\rm quad} = \frac{G_{\rm GW}}{5 \cgw c^4} \left\langle \dddot{Q}_{kl} \dddot{Q}_{mn}\right\rangle\,,
\end{align}
again recovering the result from \cite{Jimenez:2015bwa}. We emphasise that this expression holds for a time/frequency-varying $\cgw$, since it is the total power emitted at a given time/frequency, but care needs to be taken when integrating this expression to e.g. obtain the total energy emitted throughout the evolution.

We now focus on a circular inspiral and explicitly work out the power emitted in terms of the parameters describing this system. Establishing some notation, we have
\begin{align} \label{oms}
\omega_s^2 &= \frac{G_N m}{R^3}, &\omega_{\rm GW} &= 2 \omega_s, &f_{\rm GW} &= \frac{\omega_{\rm GW}}{2\pi}, &M_c &= \mu^{3/5}m^{2/5},
\end{align} 
where we recall that 
\begin{align}
\mu &\equiv \frac{m_1 m_2}{m_1+m_2},   &m &\equiv m_1 + m_2,
\end{align}
and $m_1$ and $m_2$ are the masses of the two compact objects constituting the inspiralling binary in question. 
Importantly, in \eqref{oms}, note that the expression for $\omega_s$ is effectively Kepler's law and so the appropriate gravitational coupling constant here is $G_N$, i.e. the gravitational constant that enters in the Poisson equation and which is generically different from $\Ggw$.
In this setup we can solve for $\gamma$ and find
\begin{align}
\gamma_\times(t,\theta,\phi) = \frac{4 G_{\rm GW} \mu \omega_s^2 R^2}{rc^3} \cgw \cos\theta \sin(2\omega_s t_{\rm ret}+2\phi)
\end{align}
where we note the factor of $G_{\rm GW}$ that enters following on from \eqref{solgamma} and we only show the $\times$ polarisation for brevity here -- see (3.330) and (3.331) in \cite{Maggiore:1900zz} for an analogous $+$ polarisation expression, which can be adapted along the same lines. Note that $t_{\rm ret}$ has an implicit $\cgw$ dependence as well. Now using \eqref{oms} to replace $R$ with $\omega_s$ and explicitly introducing the chirp mass, we can write
\begin{align}
\gamma_\times(t,\theta,\phi) &= \frac{G_{\rm GW}c}{G_N \cgw}\frac{4}{r}\left(\frac{G_N M_c}{c^2}\right)^{5/3}\left(\frac{\pi f_{\rm GW}}{c}\right)^{2/3} \nn \\
&\cdot \cos\theta \sin(2\omega_s t_{\rm ret}+2\phi).
\end{align}

The first fraction encodes all dark energy-related modifications we consider here, while everything that remains is as in standard GR. 
Finally, we can follow the same logic as above and compute the resulting total power emitted at any given point in time
\begin{align}
P = \frac{G_{\rm GW}c}{G_N \cgw}\frac{32 c^5}{5G_N}\left(\frac{G_N M_c \omega_{\rm GW}}{2c^3}\right)^{\frac{10}{3}}.
\end{align}

The power lost to gravitational radiation will be equal to loss of orbital energy, and, assuming Keplerian orbits, this can expressed in terms of $M_c$ and $\omega$ as~\cite{Maggiore:1900zz}
\begin{align}
P = - \frac{d E_{orbit}}{dt} = \frac{2}{3} \left( \frac{G_N^2 M_c^{5/3}}{32}\right)^{1/3} \omega^{-1/3} \dot{\omega}.
\end{align}
This can be rearranged in terms of $ \dot{\omega}$  to give
\begin{align}
\label{omegadot}
\dot\omega(t) =  \left(\frac{G_{\rm GW}c}{G_{N} \cgw}\right) \frac{3456^{1/3}}{5} \left(\frac{G_N M_c }{c^3}\right)^{5/3} \omega^{11/3}.
\end{align}
In the generic case we are considering where $\cgw$ and $G_{\rm GW}$ depend on frequency we cannot simplify further without specifying that dependency.
In the simple case that $\cgw = 1$ and $G_{\rm GW} = G_{N}$ the extra term on the right hand side vanishes and this collapses to the standard 0-PN GR solution.
In the case where $\cgw$ and $G_{\rm GW}$ deviate from the GR values but are constant over the range of observation this can be integrated analytically and rearranged
to find $h_+(t)$ and $h_{\times}(t)$. In this case we find that both the phasing evolution and the overall amplitude depend on $\cgw$ and $G_{\rm GW}$, but do so in a way
that is degenerate with chirp mass. Therefore a signal emitted in such a model would appear to be consistent with general relativity, but the recovered measurement of the chirp mass
would be offset from the true value -- also see \cite{Mirshekari:2011yq} for a related discussion of phase shifts and degeneracies. In the case of a general solution for $\cgw(\omega)$ and $G_{\rm GW}(\omega)$ this could be integrated numerically to find a model
of the emitted waveform. 

\begin{figure*}[t]
\centering
\includegraphics[width=0.99\linewidth]{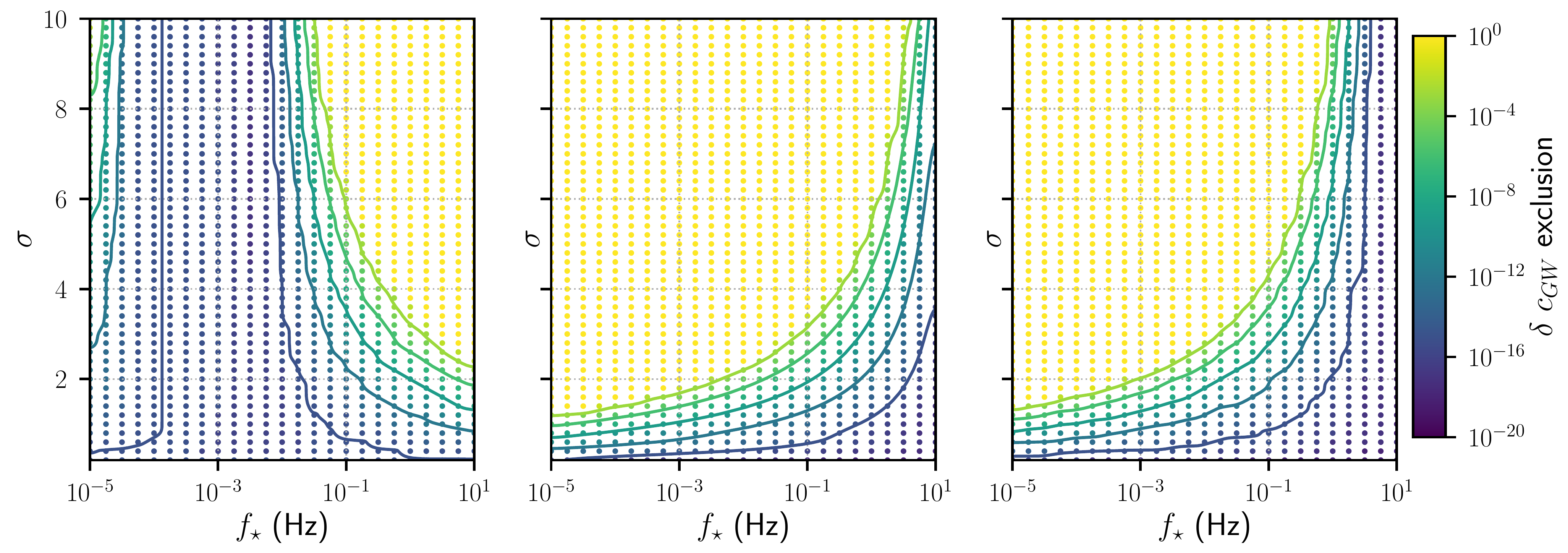}
\caption{\label{fig:exclusion_fisher_app}
The values of $\delta \cgw^{(0)}$ that would be distinguishable from a purely GR signal, as a function of the transition frequency ($f_{\star}$) and the steepness of the transition ($\alpha$). (Left) This is computed assuming a single LISA observation of a system of two supermassive black holes each having a mass of $4.154\times 10^6 M_{\odot}$ at a distance of 1Gpc. (Middle) Computed assuming an LVK observation of a GW170817-like source. (Right) Computed assuming  an ET observation of a GW170817-like source. In all cases contour lines correspond to $\delta \cgw^{(0)} = 10^{-3}, 10^{-6}, 10^{-9}, 10^{-12}$ and $10^{-15}$. 
}
\end{figure*}

In the above we have kept $G_{\rm GW}$ and $G_N$, i.e. two different gravitational coupling constants for gravitational waves and for matter, generic. However, within the context of \eqref{Horndeski_lagrangians} and inside a screened regime as is relevant for the emission of gravitational waves considered here, these are found to satisfy $G_N/G_{\rm GW} = \cgw^2/c^2$~\cite{Jimenez:2015bwa}\footnote{This relationship may break down once approaching the cutoff of the cosmological theory close to $\Lambda$, so care needs to be taken if this is to be modelled in detail.}. So here any modification of \eqref{omegadot} away from its GR limit will be controlled by a single, dimensionless parameter: $\cgw/c$. However, bounds obtained from this effect will be significantly weaker than the ones discussed in the main text. Compare the $|\delta\cgw| \lesssim 10^{-2}$ bound obtained by \cite{Jimenez:2015bwa} from the Hulse-Taylor binary using analogous reasoning with the $|\delta\cgw| \lesssim 10^{-17}$ and better constraints discussed in the main text. We therefore do not pursue bounds from this effect further here.

\section{Constraints using only LVK or ET observations} \label{app-figure-a2}

In Figure~\ref{fig:exclusion_fisher} we showed the potential constraints that could be placed on $\cgw$ when combining an
observation of a super-massive black hole merger with LISA with a
single observation of a GW170817-like source with LVK/ET. For
completeness, in Figure~\ref{fig:exclusion_fisher_app}, we also
show what the constraints would look like if \emph{only} considering
a single GW170817-like source observed with either LVK or ET, alongside
the constraints from LISA alone as comparison.

\bibliographystyle{utphys}
\bibliography{cGW_LIGO_LISA}
\end{document}